%% file: acl_latex.tex
\pdfoutput=1
\documentclass[11pt]{article}

\PassOptionsToPackage{table}{xcolor}
\usepackage[preprint]{acl}
\usepackage{enumitem}   % already in most Overleaf templates
\usepackage{amssymb}
\usepackage{fvextra}   % ← loads fancyvrb and ADDS breakanywhere, autogobble, etc.
\usepackage{enumitem}   % for customised lists
\usepackage{natbib}  
\usepackage{array}      % for \newcolumntype
\usepackage{tabularx}   % for tabularx
\usepackage{booktabs}   % if you use \toprule, \midrule, 
\usepackage{caption}
\usepackage{soul}
\usepackage{float}
\newcolumntype{L}[1]{>{\raggedright\arraybackslash}p{#1}}
\newcolumntype{C}[1]{>{\centering\arraybackslash}p{#1}}
\usepackage{booktabs}
\usepackage{siunitx}
\usepackage{placeins}
\usepackage{booktabs}    % nice horizontal rules
\usepackage{tabularx}    % automatic line-wrapping "X" column

\newcommand{\zr}{\rowcolor[HTML]{F5F5F5}}

\usepackage{multirow}   % for \multirow
\usepackage{booktabs}   % you’re using \toprule etc.
\usepackage{graphicx}   % for \resizebox (if you use it)
\DefineVerbatimEnvironment{DocChunk}{Verbatim}{
  formatcom=\footnotesize,
  breaklines=true,
  breakanywhere=true,
  breaksymbolleft={},    % nothing at the start
  breaksymbolright={},   % nothing at the end
  frame=single,
}
\usepackage{tikz}
\usetikzlibrary{arrows.meta, positioning}
\usepackage{times}
\usepackage{latexsym}
\usepackage{algorithm}
\usepackage[noend]{algpseudocode}
\usepackage[T1]{fontenc}
\usepackage[utf8]{inputenc}

\usepackage{microtype}
 \usepackage{amsmath}
\usepackage{inconsolata}

\usepackage{graphicx}

\usepackage{booktabs}
\usepackage{amsthm}     
\theoremstyle{definition}  
\newtheorem{query}{Query}
\include{notations}

\title{Respecting Temporal-Causal Consistency: Entity–Event Knowledge Graphs for Retrieval-Augmented Generation}

\author{%
  Ze Yu Zhang$^{1,2}$
  \And
  Zitao Li$^{2}$
  \And
  Yaliang Li\thanks{\;Corresponding authors.}$^{2}$
  \AND
  Bolin Ding\footnotemark[1]$^{2}$
  \And
  Bryan Kian Hsiang Low\footnotemark[1]$^{1}$
  \AND
  \normalfont
  $^{1}$Department of Computer Science, National University of Singapore \\
  $^{2}$Alibaba Group\\
  \texttt{\{zhan1130,lowkh\}@comp.nus.edu.sg} \\
  \texttt{\{zitao.l,yaliang.li,bolin.ding\}@alibaba-inc.com}
}
\begin{document}
\maketitle
\begin{abstract}
% In this work, we propose entity-event KG RAG, a novel KG RAG approach that is more effective for RAG tasks that involve sophisticated understanding of text...

% In addition, we also introduce a new dataset, ..., for benchmarking RAG methods on such tasks.

Retrieval-augmented generation (RAG) based on large language models often falters on narrative documents with inherent temporal structures. Standard unstructured RAG methods rely solely on embedding-similarity matching and lack any general mechanism to encode or exploit chronological information, while knowledge graph RAG (KG-RAG) frameworks collapse every mention of an entity into a single node, erasing the evolving context that drives many queries. To formalize this challenge and draw the community’s attention, we construct \benchmark, a robust and discriminative QA benchmark that measures temporal, causal, and character consistency understanding in narrative documents (e.g., novels) under the RAG setting. We then introduce \textit{Entity-Event RAG} (\ours), a dual-graph framework that keeps separate \emph{entity} and \emph{event} subgraphs \emph{linked by a bipartite mapping}, thereby preserving the temporal and causal facets needed for fine-grained reasoning. Across \benchmark, our approach outperforms state-of-the-art unstructured and KG-based RAG baselines, with notable gains on causal and character consistency queries. \ours therefore offers a practical path to more context-aware retrieval for tasks that require precise answers grounded in chronological information.
% \zt{make connection of chronological to causal/temporal consistency somewhere in the intro}.
% without requiring a longer context window or model finetuning.
\end{abstract}

\section{Introduction}

Large language models (LLMs) have demonstrated remarkable zero‑shot and few‑shot capabilities across various NLP tasks. 
Yet, LLMs remain fundamentally constrained by their fixed context window: once the prompt exceeds a certain length, processing becomes slow and the model’s reasoning capability is significantly impaired \citep{liu2023lost, fictionlivebench2025,li2024long, gao2024insights,an2024does}. To mitigate this limitation, retrieval‑augmented generation (RAG) was introduced, coupling the generator with an external document retriever so that fresh, grounded evidence can be injected on demand \citep{lewis2020retrieval}.  RAG not only extends an LLM's effective memory but also reduces hallucinations and allows rapid knowledge updates without costly re‑training \citep{shuster2021retrieval,lewis2020retrieval,borgeaud2022improving,bechard2024reducing}.

Vanilla unstructured RAG, however, treats each passage in isolation and therefore struggles
with reasoning that depends on a document's temporal or causal structure. For example, the following seemingly straightforward question for \textit{Harry Potter} enthusiasts can be challenging for a RAG system to answer reliably:
% For example, in the novel \textit{Harry Potter and the Sorcerer's Stone},
% consider the question:

% \zt{use format as query1}
\begin{query}
\emph{Who was jinxing Harry's broom during his first Quidditch match?}
\end{query}

Because effects in a story typically follow their causes in time, losing chronological ordering also obscures causal links. If a RAG system naively retrieves the seemingly most relevant passage from the earlier part of the story, it will incorrectly conclude that \emph{Snape} was responsible for hexing Harry's broom. This error arises because it overlooks crucial information revealed later—that \emph{Professor Quirrell} was actually causing the interference, while Snape's suspicious actions were attempts to counteract Quirrell's jinx. The key reason is that the document's \emph{chronological context information}, which is crucial for generating answers that are both temporally coherent and causally sound, is not preserved in the vanilla unstructured RAG after being chunked.

To capture richer structure, knowledge‑graph (KG) RAG methods such as GraphRAG build an entity graph over the corpus and retrieve communities of related nodes \citep{edge2024local}.  
Unfortunately, the standard graph‑construction pipeline relies on aggressive de‑duplication of entity mentions; this collapses distinct temporal or contextual facets of the same character into a single node, discarding information that is crucial in domains like narrative fiction, where characters evolve continuously. Consider the following query:
% Hermione Granger is introduced as a rule‑obsessed know‑it‑all who reprimands classmates for the smallest infractions (Ch.~9, "The Midnight Duel"), yet after the troll incident (Ch.~10) she forges a deep friendship with Harry and Ron and is soon helping them brew an illegal potion and sneak past teachers to protect the Stone.  If every ``Hermione Granger'' mention is collapsed into a single KG node, a query
% such as

\begin{query}\label{q:hermione}
\emph{How would Hermione Granger react if a friend proposed breaking school rules after the troll incident?}
\end{query}

Hermione is introduced as a rule‑obsessed know‑it‑all who reprimands classmates for the smallest infractions (Ch.~9), yet after the troll incident (Ch.~10), she forges a deep friendship with Harry and Ron and is soon helping them brew an illegal potion and sneak past teachers to protect the Stone.  
If every ``Hermione Granger'' mention is collapsed into a single KG node, a query will likely retrieve only her predominant rule‑abiding persona—the class monitor who scolds Ron for casting \emph{Lumos} in the corridor—while overlooking
later chapters in which she calmly steals ingredients and slips past teachers to protect the Stone (Ch.~16). 
Thus, the aggressive de‑duplication of entity mentions in KG-RAG methods can erase her arc from strict disciplinarian to pragmatic rule‑breaker, masking precisely the nuance the question probes and leading the system to an outdated, inaccurate answer.

To rigorously define the challenge of temporal-causal consistency in RAG applications and evaluate how current RAG systems handle the aforementioned narrative-focused challenges, we construct \textbf{\benchmark}, a retrieval-style QA benchmark drawn from nine public-domain  narrative documents.  Every question targets one of eight fine-grained reasoning facets, ranging from causal and character consistency to symbolism and thematic insight.

 To address the aforementioned challenges, we propose an \emph{Entity–Event KG} composed of (i) an entity subgraph; (ii) an event subgraph; and
(iii) a \emph{bipartite edge set} that maps each entity mention to the
events in which it participates, thereby preserving those nuances. Instead of collapsing duplicates, we first extract both entities \emph{and} their associated events, then link every event to the \emph{specific} mention of each participating entity; because we never merge mentions that arise in different parts of the story, each entity node carries its own \emph{context-specific description}. The resulting graph retains multiple, context‑specific representations of entities while still exposing the relational structure needed for retrieval. We term the resulting RAG framework, which retrieves over the proposed entity–event KG, \textit{Entity–Event RAG} (\ours). At retrieval time, beyond the vanilla pipeline, we introduce an augmentation operation that calls the backbone LLM only once to inject richer context. Empirically, this single-call augmentation lets \ours achieve significant gains on fine-grained narrative-reasoning queries, outperforming state-of-the-art unstructured and KG-based RAG baselines and pointing to a practical path toward more context-aware retrieval for LLMs.

The contributions of this work are as follows:

    \noindent$\bullet$ \textit{ We release \benchmark, the first open, passage-grounded benchmark that stresses temporal and causal reasoning over book-length narratives under a RAG setting.}
    
    \noindent$\bullet$ \textit{ We propose \ours, a dual-graph retrieval framework that keeps every entity mention distinct and grounded in its associated event, thereby preserving the causal and temporal consistency of evolving characters and plot.}
    
    \noindent$\bullet$ \textit{ Extensive experiments demonstrate that variants of \ours outperform state-of-the-art unstructured and KG-based RAG baselines on \benchmark.}

\section{Related Work}
\subsection{Retrieval-Augmented Generation (RAG)}
Although today's large language models can store vast amounts of world knowledge, that knowledge is {\em static}, {\em unverifiable}, and {\em expensive} to refresh.  \citet{lewis2020retrieval} demonstrated that attaching a live retriever to a generator lets the model pull {\em up‑to‑date evidence on demand}, achieving higher accuracy than closed‑book BART while returning the very passages that support each claim. Subsequent analyses show that retrieval‑augmented language models markedly cut hallucination rates—especially on fact‑heavy tasks—because retrieved text provides an external "ground truth" that the decoder can copy or paraphrase instead of guessing \citep{lv2024coarse,sree2025hybrid}. \citet{gao2023precise} go a step further: their \textsc{HyDE} method lets the LLM hallucinate a "hypothetical" answer, embeds it, and uses that
vector to query the index, delivering \emph{precise zero-shot dense retrieval} without relevance labels—an idea we later adapt in Section~\ref{sec:methodology}. RAG also sidesteps model‑update costs: keeping the index current is far cheaper and faster than re‑training or fine‑tuning multi‑billion‑parameter networks, a point underscored by adaptive systems such as Self‑RAG and Speculative RAG that retrieve only when necessary and verify drafts to stay both efficient and factual \citep{lv2024coarse,lewis2020retrieval}.  Finally, long‑context methods like LongRAG show that retrieval can extend an LLM's effective memory without quadratic attention overhead, enabling faithful reasoning over book‑length evidence while keeping latency low\citep{jiang2024longrag}.   In short, RAG equips LLMs with a dynamic, interpretable and cost‑effective memory, tackling three core limitations—knowledge staleness, hallucination, and context length—that purely parametric models struggle to overcome.

\subsection{Knowledge Graph based RAG}

GraphRAG \citep{edge2024local} explicitly builds an entity‐level knowledge graph, run community detection, and then retrieve and summarize the most query‑relevant communities, thereby improving the relevance of the top‑$k$ chunks with respect to a query.  Follow‑up studies highlight three practical drawbacks of this design: (i) the multi‑pass entity/relation extraction and community‑summary generation make preprocessing costly in both tokens and compute, (ii) traversing and summarising the graph at inference time adds 2–3× higher end‑to‑end latency, and (iii) the graph index and its summaries grow super‑linearly with corpus size, complicating incremental updates and ballooning memory usage \citep{wang2025archrag,chen2025pathrag,peng2024graph}.

LightRAG~\citep{guo2024lightrag} tackles these limitations by folding relational signals into a standard dense index and introducing a dual‑level, coarse‑to‑fine retriever that first selects cluster representatives and then expands to their ego networks; this removes explicit graph traversal, supports incremental index patches, and reduces indexing token cost by $\approx 60\%$ while roughly halving median query latency, all without hurting answer quality on UltraDomain, QFS and other multi‑hop QA benchmarks\citep{guo2024lightrag,chen2025pathrag}.

\subsection{Long-context narrative benchmarks}
Recent work has begun to probe whether LLMs can reason over book-length
inputs without truncation.  \textsc{XL$^2$Bench}~\citep{ni2024xl},
\textsc{Loong}~\citep{wang2024leave}, and \textsc{LongGenBench}%
~\citep{liu2024longgenbench} extend QA or generation tasks to 100 K-token
contexts, while the very recent \textsc{Fiction.LiveBench}%
~\citep{fictionlivebench2025} packages full user-written stories into the prompt and asks multi-step comprehension
questions that require tracking characters and foreshadowing across tens of
thousands of tokens.  
Because every benchmark above \textit{gives the model the entire story up front},
they measure intrinsic long-context reasoning rather than the retrieval quality.

\section{New Benchmark: \benchmark}
\label{sec:Dataset}

 % \paragraph{Why a new benchmark?}
% The datasets above reveal how far LLMs can \emph{read} once the entire book or
% document is placed in their prompt, yet they leave a few practical
% questions unanswered for RAG-based approach:  
% \textbf{(i)} \emph{Can a RAG system retrieve the right snippets when
% the full story does \emph{not} fit into context?} and  
% \textbf{(ii)} \emph{Can it reason over a character's time-specific state—e.g.\ "late-story Hermione" versus "early-story Hermione"—without that
% nuance being lost to entity de-duplication?} Moreover, \textbf{(iii)} existing benchmarks typically provide \emph{no} passage-level evidence for their ground-truth answers; and \textbf{(iv)} many rely on fully private evaluation framework with no publicly available data. These existing issues makes it difficult for researchers to verify reported results or to probe where and why current RAG systems fail.

Although the datasets mentioned above for long context reasoning tasks can be adapted to benchmarking the RAG methods, there are very limited resources that can be used to evaluate an RAG method specifically on the temporal-causal consistency question: 
\textbf{(i)} Most existing datasets focus on general long context tasks, but have limited focus on these challenging reasoning tasks related to temporal-causal consistency.
For example, a rare existing task requires reasoning over a character's time-specific state (e.g., "late-story Hermione" versus "early-story Hermione").
\textbf{(ii)} Existing benchmarks typically provide \emph{no} passage-level evidence for their ground-truth answers.
Different from the long context tasks, which only evaluate the model answers, evaluating the retrieval snippets is also a common metric when benchmarking the RAG systems. 
However, without the passage-level evidence, extra effort may be required to determine the retrieval stage's correctness.
\textbf{(iii)} Although some datasets (e.g., \citet{fictionlivebench2025}) also focus tasks based on narrative documents, they rely on a entirely private evaluation framework with no publicly available data. These existing issues make it difficult for researchers to verify reported results or to probe where and why current RAG systems fail.

% To test different RAG systems' capability of 
% reasoning with respect to the temporal and causal structure of the document, we propose a new QA dataset that asks questions about ...  for 9 different stories, ranging across novels, musical scripts and children stories. The questions can be broadly categorized under the following types:
% "Causal Consistency",
%     "Character and Behavioural Consistency",
%     "Setting, Environment \& Atmosphere",
%     "Symbolism, Imagery and Motifs",
%     "Thematic, Philosophical \& Moral",
%     "Narrative \& Plot Structure",
%     "Social, Cultural and Political",
%     "Emotional and Psychological".

% Here are some examples:

% All the query are generated by an oracle LLM. We choose o1-pro and claude-3.7-sonnet as the oracle LLMs, because they are the best model for such task as the time we started the experiment. For each query, the oracle model also provides a ground truth.

To test whether different RAG systems can reason over a document's {\em temporal} and {\em causal consistency}, we introduce \textbf{\benchmark}—a  QA benchmark built from nine narrative works spanning novels, musical scripts and children's stories.\footnote{The underlying literary works are in the public domain in the United States.  
The e-book files were obtained from Project~Gutenberg and are redistributed under the
\emph{Project~Gutenberg-tm License} that accompanies each file.}  For each story we automatically generate questions that probe eight fine-grained reasoning facets:

% \begin{itemize}
% [leftmargin=0.65cm,labelsep=0.2cm,font=\normalfont\itshape]

\noindent$\bullet$ \textit{Causal Consistency} – cause and effect, logical sequences, or explanations of how events unfolded.  

\noindent$\bullet$ \textit{Character \& Behavioural Consistency} – character motivations, development, or psychology.  

\noindent$\bullet$  \textit{Setting, Environment \& Atmosphere} – physical locations, time periods, or mood/atmosphere

\noindent$\bullet$ \textit{Symbolism, Imagery \& Motifs} – symbolic elements, recurring imagery, or metaphorical representations. 

\noindent$\bullet$ \textit{Thematic, Philosophical \& Moral} – deeper meanings, philosophical ideas, or ethical implications

\noindent$\bullet$ \textit{Narrative \& Plot Structure} – story organization, plot devices, or narrative techniques.

\noindent$\bullet$ \textit{Social, Cultural \& Political} – societal contexts, cultural elements, or political dimensions.

\noindent$\bullet$ \textit{Emotional \& Psychological} – emotional responses, psychological states, or mental processes\@.  
% \end{itemize}

The questions from each category require solid understanding grounded on the progression of the story and cause-and-effect of the elements involved in the plot. Our \textbf{\benchmark} closes the aforementioned gap with:

  \noindent$\bullet$ \textbf{Focus on temporal and causal consistency} — each query hinges on a character's \emph{specific temporal facet} (e.g.\ "after the troll incident"), so retrieving an early-story snippet fails.
  
    \noindent$\bullet$  \textbf{Passage-level supervision} — every answer is paired with exact start/end byte offsets, making verification straightforward and reliable.
    
    \noindent$\bullet$  \textbf{Unrestricted accessibility} — every narrative is drawn from Project Gutenberg, putting the full texts in the public domain. Unlike fully private benchmark, This guarantees that anyone can obtain, inspect, and redistribute the benchmark data without restrictions.

Thus ChronoQA tests whether a RAG system can (1) fetch the right snippet when the full story cannot fit into context and (2) reason over evolving entity states that standard KG de-duplication erases.

% Queries are produced with GPT-o1-pro and Claude-3.7-Sonnet acting as "oracle" LLMs; for each question the oracle also returns an answer that is used as ground truth.

\paragraph{Dataset statistics.}  The final release contains \textit{9} stories, \textit{497} question–answer pairs. The actual number of question–answer pairs for each story and categories can be found in Table~\ref{tab:per_document} and \ref{tab:per_category}. Other details can be found in Appendix~\ref{appendix:Statistics}.

\addcontentsline{toc}{section}{Appendix A. chronoQA Statistics}

\begin{table}[h]
\centering
\scriptsize
\caption{Questions per story (total = 497).}
\begin{tabular}{lr}
\toprule
Story & \# Questions \\
\midrule
The Wonderful Wizard of Oz               & 82 \\
The Hound of the Baskervilles            & 72 \\
The Phantom of the Opera                 & 70 \\
A Study in Scarlet                       & 67 \\
The Sign of the Four                     & 62 \\
Harry Potter and the Chamber of Secrets                       & 55 \\
The Adventures of Sherlock Holmes        & 34 \\
Harry Potter and the Sorcerer's Stone    & 30 \\
Les Misérables                           & 25 \\
\midrule
\textbf{Total}                           & \textbf{497} \\
\bottomrule
\end{tabular}
\label{tab:per_document}
\end{table}

\begin{table}[h]
\centering\scriptsize
\caption{Questions per reasoning category (total = 497).}
\begin{tabular}{lr}
\toprule
Category & \# Questions \\
\midrule
Character Consistency                    & 229 \\
Causal Consistency                       &  96 \\
Symbolism, Imagery \& Motifs             &  56 \\
Thematic, Philosophical \& Moral         &  36 \\
Narrative \& Plot Structure              &  31 \\
Setting, Environment \& Atmosphere       &  25 \\
Social, Cultural \& Political            &  22 \\
Emotional \& Psychological               &   2 \\
\midrule
\textbf{Total}                           & \textbf{497} \\
\bottomrule
\end{tabular}
\label{tab:per_category}
\end{table}

\paragraph{Generation pipeline.}
Because every story is short enough to fit within the
oracle's context window, we do not chunk the text.  Instead we use a
\emph{two-stage} pipeline:

\noindent$\bullet$ \textbf{Q–A candidates generation:}  
        The entire story is provided to the oracle model (GPT-o1-pro \citep{jaech2024openai} and Claude-3.7-Sonnet \citep{anthropic2025claude37}) together with a
        category list and an instruction to propose up to diverse
        question–answer pairs per category (prompt template in
        Appendix~\ref{subsec:DataGenerationPrompts}).  The model's answer is
        retained as provisional ground truth. 

\noindent$\bullet$ \textbf{Verification, Filtering and de-duplication:}    For each ground truth, the oracle model is requried to output the starting sentence and ending sentence of the excerpt where the answer can be inferred. Together with the starting sentence and the ending sentence, we also provide the excerpt and its exact start/end byte offsets in the dataset.   We discard candidate questions that are duplicates or for which the excerpt extracted does not substantiate the answer.

\paragraph{Dataset release.} ChronoQA is accessible via Hugging Face: \url{https://huggingface.co/datasets/zy113/ChronoQA}.

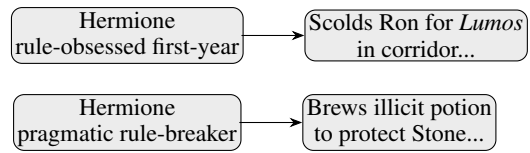
\begin{figure}[t] % [t], [h], etc. as you prefer
  \centering
\tikzset{
  every node/.style = {inner sep=1.5pt, font=\small, rounded corners,
                       draw, fill=gray!15, align=center}, % <- align added
  flush arrow/.style = {->, shorten <= -.8pt}
}
\begin{tikzpicture}[>=Stealth, node distance={4mm and 8mm}]
  \node (e1) {Hermione\\rule-obsessed first-year};
  \node[right=8mm of e1] (ev1)
        {Scolds Ron for \textit{Lumos}\\in corridor...};

  \node[below=4mm of e1] (e2)
        {Hermione\\pragmatic rule-breaker};
  \node[right=8mm of e2] (ev2)
        {Brews illicit potion\\to protect Stone...};

  \draw[flush arrow] (e1.east) -- (ev1.west);
  \draw[flush arrow] (e2.east) -- (ev2.west);
\end{tikzpicture}
  \caption{Illustration of two temporally distinct Hermione mentions and their associated events.}
  \label{fig:entity-event-demo}
\end{figure}

\begin{figure*}[!ht]
    \centering
    \includegraphics[width=0.8\linewidth]{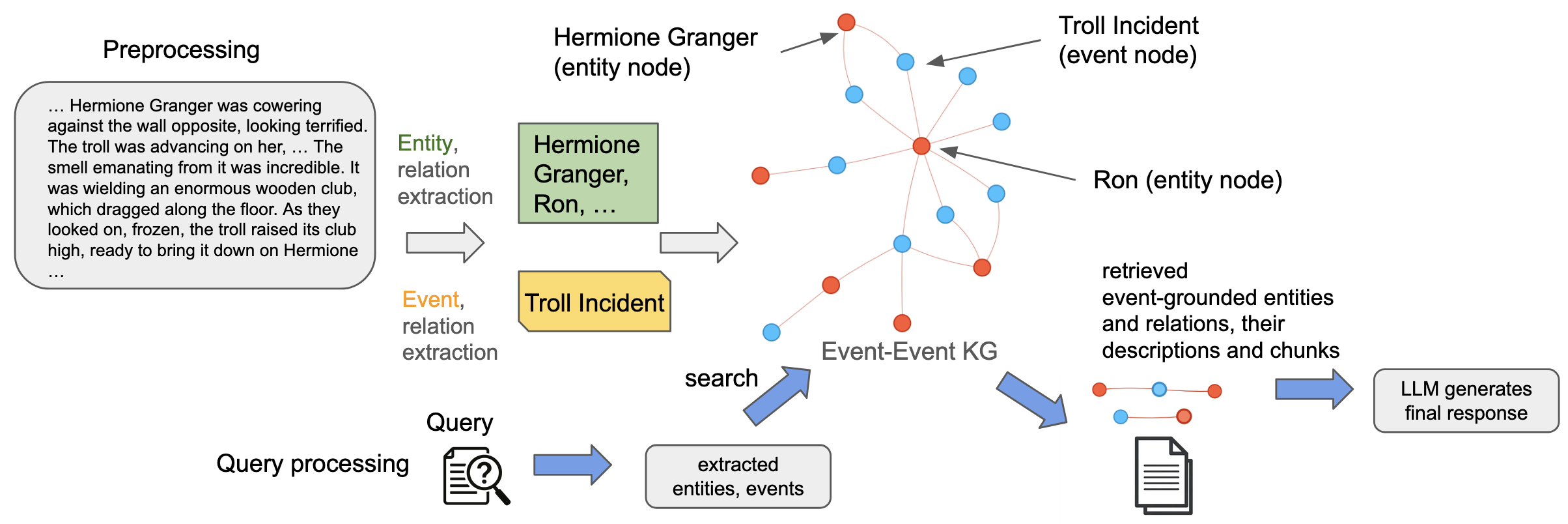}
    \caption{
        Overall architecture of the \ours framework.
    }
    \label{fig:architecture}
\end{figure*}

\section{Methodology}
\label{sec:methodology}

Standard KG-RAG preprocessing merges every mention of entities extracted from chunked document (e.g., \emph{Hermione Granger},
\emph{Professor Quirrell}, or the \emph{Sorcerer's Stone}) into a single
node—obliterating the time-specific information. Such nuance is needed to answer questions such 
as Query~\ref{q:hermione}. We therefore keep \textbf{each} mention---a concrete, context-specific instance of an entity as it appears in a single chunk of text (e.g., "Hermione" in Chapter 9 versus "Miss Granger" in Chapter 16) --- distinct and tether it to the \emph{event
snippet} in which it occurs. Intuitively, the structure is as shown in Figure~\ref{fig:entity-event-demo}. The left column holds \textit{entity mentions} and their contextual descriptions; the right column holds \textit{events} and their descriptions.
Treating the entity as two disjoint vertex sets ensures temporal facets never collapse.

\paragraph{Formal definition.}
Let  
\(
  \mathcal{G}_{\text{ent}}
     =(V_{\text{ent}},E_{\text{ent}})
\)
be the directed graph of entity mentions and  
\(
  \mathcal{G}_{\text{evt}}
     =(V_{\text{evt}},E_{\text{evt}})
\)
the graph of events.  
We connect the two with a bipartite edge set
% \begin{equation}
% \mathcal{B} \;=\;
% \bigl\{(v_i,e_j)\;\big|\;
%       v_i\in V_{\text{ent}},~
%       e_j\in V_{\text{evt}},~
%       \mathrm{Name}(v_i)\subset\mathrm{Desc}(e_j)\bigr\},
% \label{eq:bipartite}
% \end{equation}

\begin{align}
\mathcal{B}
  = \bigl\{(v_i,e_j)\,\big|{}&
      \;v_i \in V_{\text{ent}},\;
      e_j \in V_{\text{evt}},\notag\\[2pt]
  & \mathrm{Name}(v_i)\subset\mathrm{Desc}(e_j)
     \bigr\}.                       \label{eq:bipartite}
\end{align}

so that an edge exists if and only if the entity's name $\mathrm{Name}(v_i)$ appears in the event
description $\mathrm{Desc}(e_j)$.

% The triple
% \(
%   (\mathcal{G}_{\text{ent}},\mathcal{G}_{\text{evt}},\mathcal{B})
% \)
% is the backbone of our retrieval pipeline and will re-appear in Alg.~\ref{alg:retrieval}. 

\paragraph{Preprocessing.} 
% The preprocessing mirrors Algs.~\ref{alg:insertion} and~\ref{alg:bipartite}, and it unfolds in four steps:
\ours consists of the following preprocessing steps to build knowledge graphs\footnote{Pseudocode can be found as Algs.~\ref{alg:insertion} and~\ref{alg:bipartite} in the appendix.}.

\vspace{0.5em}

\noindent \textit{1. Chunking.} 
      The document is split into chunks
      \(\{c_\ell\}_{\ell=1}^{L}\) and stored in a key–value store.

% \item \textbf{Creating entity and event sets.}  
%       Each chunk \(c_\ell\) is fed twice to an LLM—
%       once with an \emph{entity} prompt and once with an \emph{event} prompt.
\vspace{0.5em}
      \noindent \textit{2. Creating entity and event sets.}  
       Each chunk \(c_\ell\) is fed twice to an LLM to extract \emph{entities} $V_{\text{ent}}^{(\ell)}$ and \emph{events} $V_{\text{evt}}^{(\ell)}$ respectively.
       % —once with an \emph{entity}
       % prompt and once with an \emph{event} prompt to undergo extraction. 
       For every entity and event extracted, we also ask the LLM for a one-sentence description. Note that in entity's case, this description will be \emph{context-aware} so the same entity extracted from different document chunks will have different descriptions\footnote{e.g.\ "Hermione—rule-obsessed first-year" versus "Hermione—pragmatic rule-breaker".}.
       This description is stored in the node's \texttt{description} field and is kept
       \emph{distinct} for every mention. The extractions form the node sets  
      \(V_{\text{ent}}=\!\bigcup_\ell V_{\text{ent}}^{(\ell)}\) and  
      \(V_{\text{evt}}=\!\bigcup_\ell V_{\text{evt}}^{(\ell)}\).
      Co-mentions inside the same chunk yield the directed edge sets
      \(E_{\text{ent}}\) and \(E_{\text{evt}}\) exactly as in LightRAG.
      
\vspace{0.5em}

\noindent \textit{3. Link entities to events (\(\mathcal{B}\)).}  
      For every entity \(v\!\in\!V_{\text{ent}}^{(\ell)}\) we scan the events
      \(e\!\in\!V_{\text{evt}}^{(\ell)}\);  
      if the node's name \(\mathrm{Name}(v)\) appears in
      \(\mathrm{Desc}(e)\) we add the bipartite edge \((v,e)\) to
      \(\mathcal{B}\) (Eq.~\ref{eq:bipartite}).  
      
\vspace{0.5em}

\noindent \textit{4. Embed and index.}  
      Each node's name and
      description is embedded with embedding model \(f(\cdot)\) and written to two
      vector stores—one for entities, one for events—while the full graph
      \(\mathcal{G}_{\text{ent}}\cup\mathcal{G}_{\text{evt}}\cup\mathcal{B}\)
      is saved for hop-limited traversals.

Figure~\ref{fig:architecture} shows the overall pipeline of \ours and Figure~\ref{fig:e2-KG} shows a subgraph of the Entity-Event KG of \textit{Harry Potter and the Sorcerer's Stone} and an event node's meta data.

\paragraph{Query-time retrieval.}
\label{sec:retrieval}
Given a query $q$, the retrieval mechanism conducts the following steps.
% (or a HyDE variant; see \ref{sec:hyde-coupling}),
% we perform:

\vspace{0.5em}
\noindent \textit{1. Cue extraction.}  
% The entity and event LLM extractor maps \(q\) to phrase sets \(S_{\text{ent}},S_{\text{evt}}\).
A pair of entity and event phrase sets are first extracted by an LLM extractor $g(\cdot)$, namely $(S_{\text{ent}},S_{\text{evt}}) = g(q)$

\vspace{0.5em}
\noindent \textit{2. Embedding generation.} 
The $(S_{\text{ent}},S_{\text{evt}})$ will be fed into the embedding model $f(\cdot)$ and generate a embedding $z$ used for retrieval, namely $z = f(g(q))$.

\vspace{0.5em}
\noindent \textit{3. Seed nodes retrieval.} 
With the embedding $z$, the next step is to retrieve seed nodes vector stores \(V_q\subseteq V_{\text{ent}}\cup V_{\text{evt}}\).

\vspace{0.5em}
\noindent \textit{4. One-hop expansion.}  
We take one step over the bipartite edges,
\(V_q^{+}=V_q\cup\{e\mid\exists v\in V_q:(v,e)\in\mathcal{B}\}
             \cup\{v\mid\exists e\in V_q:(v,e)\in\mathcal{B}\}\),
guaranteeing that every retrieved entity comes packaged with its
time-specific events and vice-versa.

\vspace{0.5em}
\noindent \textit{5. Similarity ranking.}
All passages are embedded offline; at query time we select only the vectors
of nodes in $V_q^{+}$—typically top-$k$ based on their similarity instead of the full index. 
When multiple entity nodes correspond to the \emph{same}
entity, their similarity scores are re-ranked according to the similarity
ranking of their \emph{associated event} nodes.  This grounds each entity in the
events’ evidence, preserving temporal–causal consistency.

\vspace{0.5em}
\noindent \textit{6. Context assembly.}  
We collect  
(i) the raw passages behind the top-\(k\) nodes;  
(ii) a linearised dump of the subgraph
\(\mathcal{G}_{\text{sub}}
    =(V_q^{+},
      (E_{\text{ent}}\!\cup\!E_{\text{evt}}\!\cup\!\mathcal{B})
      \cap (V_q^{+}\!\times\!V_q^{+}))\); 
and  
(iii) the metadata (descriptions, labels, etc.) associated
with every selected node and edge. This enriched, structured context is fed to the backbone LLM, allowing it to answer questions that demand fine-grained temporal and causal reasoning
using the truly relevant document chunks.

Together, these two routines turn the intuition from Figure~\ref{fig:entity-event-demo} into a fully operational retrieval pipeline, with \(\mathcal{B}\) acting as the
critical "glue’’ that preserves evolving entity states across time.

\label{sec:hyde-coupling}
\paragraph{Hypothetical response coupling.} 

One potential limitation of \ours is that the retrieval effectiveness heavily depends on how much information can be extracted from the query. If the query lacks details, the extraction might not be able to capture enough information, preventing the core mechanism from being fully utilized. To make the \emph{entity--event} approach more effective, we incorporate the core idea in
HyDE~\citep{gao2023precise}.
HyDE first asks the backbone LLM to draft a \emph{hypothetical response} without the document,
then merges the embedding of the hypothetical response with the original query before performing similarity-based top-$k$ retrieval. Even though the content might be factually off, the inclusion of hypothetical response provides a richer context for similarity matching.

% We couple HyDE with our entity--event pipeline in four distinct ways:
We introduce four variants of hypothetical response mechanism in \ours for retrieving more accurate and comprehensive information. To clearly present the them, we use $h$ denote hypothetical response and $[q;h]$ as text-level concatenation of the original query $q$ and hypothetical response $h$.

\vspace{0.5em}
\noindent \textit{1. Combined extraction (Comb. extraction).}  
Instead of providing the the query $q$ to the extractor, this approach gives $[q;h]$ as the input to the extractor. 
Thus, the generated embedding can be represented as $z=f(g[q;h])$.

\noindent \textit{2. Hypothetical extraction (Hyp. extraction)  .} 
Similarly, this variance replaces the original query $q$ with the hypothetical answer $h$ to the extractor, and use the output to generate embeddings, i.e., $z=f(g(h))$.

\noindent \textit{3. Combined embedding (Comb. embedding).}  
This method omits the extractor, and directly embeds the concatenated the hypothetical answer and the original query, i.e., $f([q;h])$.

\noindent \textit{4. Hypothetical embedding (Hype. embedding).}  
Similar to the above one in terms of removing the extractor step, we directly generate embeddings with the hypothetical documents, i.e., $z=f(h)$.

We experiment with these four variants to investigate the importance of two key factors: incorporating the original query and employing the extraction step. Specifically, we examine how the presence or absence of the original query and the extraction mechanism affect retrieval effectiveness. After generating embeddings using each approach, the following steps are identical to the query-time retrieval steps 3 to 6 introduced earlier.

\begin{table}[!b]
\centering
\scriptsize
\caption{Overall average scores}
\label{tab:overall-avg-scores}

% zebra-striping
\rowcolors{3}{gray!10}{white}

\begin{tabular}{
  @{}r                         % Rank
  l                            % Mode
  S[table-format=1.4]          % Avg Score (4 dp)
  S[table-format=5,            % Total—no decimals
    round-precision=0]@{}}     
\toprule
\textbf{Rank} & \textbf{Mode} & \textbf{Avg Score} & \textbf{Total}\\
\midrule
\rowcolor{yellow!20} 1 & \ourss (comb. extraction)      & 7.1257 & 10603 \\
\rowcolor{yellow!20} 2 & \ourss (comb. embedding)       & 7.0719 & 10523 \\
% \rowcolor{yellow!20} 3 & \ourss (vanilla)\_averaged\_embedding       & 7.0551 & 10498 \\
\rowcolor{yellow!20}  3  & \ourss (hyp. extraction)  & 6.9832 & 10391 \\
4  & \ourss (hyp. embedding)   & 6.9395 & 10326 \\
5  & LightRAG hybrid                                         & 6.8804 & 10238 \\
6  & GraphRAG drift                                & 6.8206 & 10149 \\
7  & GraphRAG local                                & 6.7997 & 10118 \\
8  & \ourss (vanilla)                                  & 6.7083 &  9982 \\
9 & vanilla RAG                                          & 6.6022 &  9824 \\
10 & LightRAG local                                          & 6.5497 &  9746 \\
11 & GraphRAG global                               & 6.5087 &  9685 \\
12 & LightRAG global                                         & 6.4583 &  9610 \\
13 & vanilla HyDE                                 & 6.3555 &  9457 \\
14 & LightRAG hybrid (comb. embedding)              & 5.6969 &  8477 \\
15 & LightRAG hybrid (comb. extraction)             & 5.6815 &  8454 \\
16 & LightRAG hybrid (hyp. embedding)          & 5.6714 &  8439 \\
17 & LightRAG hybrid (hyp. extraction)         & 5.6552 &  8415 \\
18 & RQ-RAG                                          & 3.5141 &  5229 \\
\bottomrule
\end{tabular}
\end{table}

\begin{figure*}
\centering
\begin{minipage}{0.3\textwidth}
    \centering
    \captionof{table}{Preprocessing runtimes for the three KG RAG systems}
  \label{tab:pipeline-times}

    \resizebox{\linewidth}{!}{
        \centering
        \begin{tabular}{@{}l l r@{}}
    \toprule
    System & Stage & Time (s) \\ 
    \midrule
    LightRAG & Total pipeline & 101.1723 \\ 
    \midrule
    \multirow{4}{*}{\ourss}
      & Entity subgraph construction   & 108.3549 \\ 
      & Event subgraph construction    & 103.6091 \\ 
      & Bipartite-mapping       &   0.5076 \\
      & \textbf{Total pipeline}        & 108.8625\\
    \midrule
    GraphRAG & Total pipeline & 208.9904 \\ 
    \bottomrule
  \end{tabular}
    }
\end{minipage}\hfill%
\begin{minipage}{0.7\textwidth}
    \centering
    \includegraphics[width=1.0\linewidth]{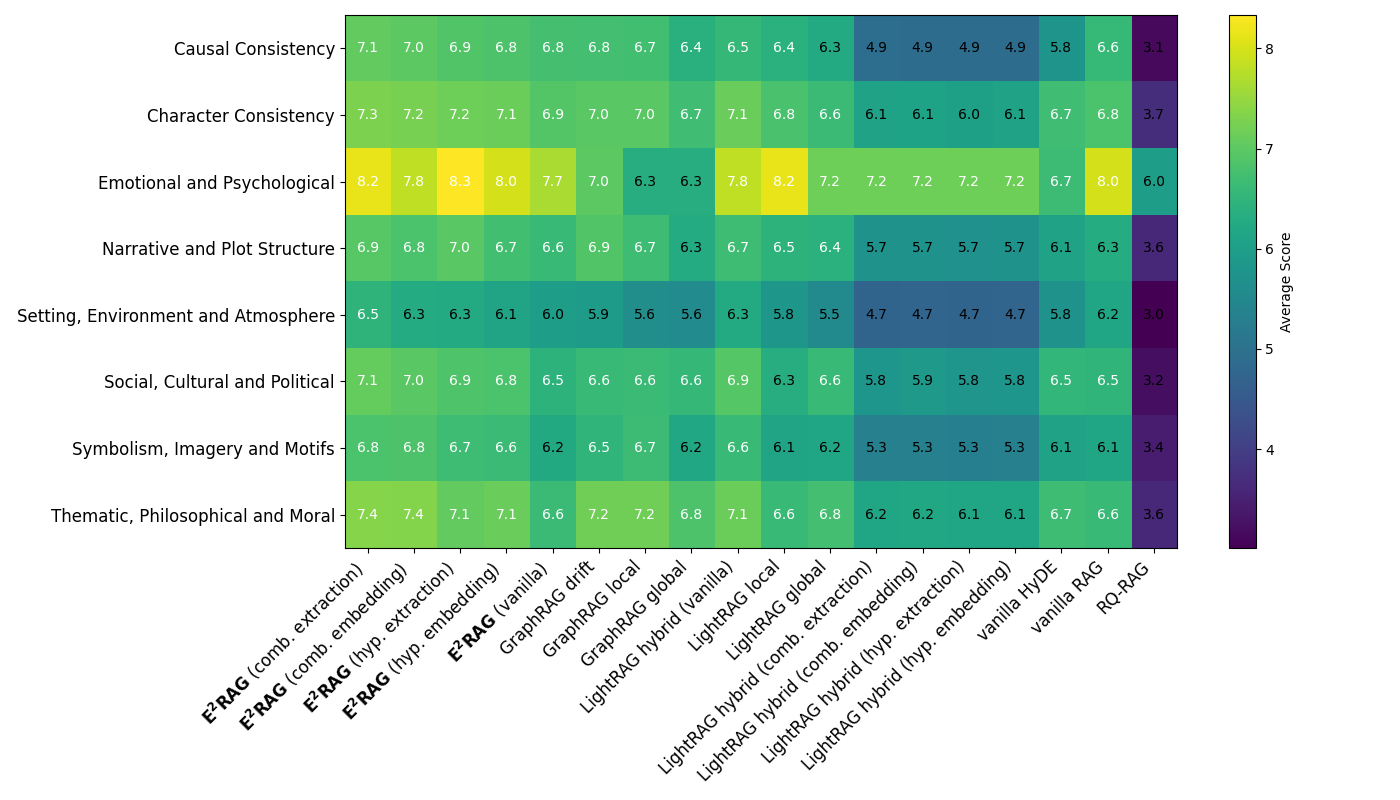}
    \caption{
        Heatmap of the scoring for each category and mode.
    }
    \label{fig:heatmap}
\end{minipage}
\end{figure*}

\section{Experiments}
\label{sec:Experiments}
% In the experiment, we use different modes of LightRAG and GraphRAG, HyDE, RQ-RAG and vanilla RAG as baselines. To evaluate the performance of each method, for each query, we give an evaluator LLM the query, the ground truth and the responses of the variant of \ours and all baselines. The evaluator LLM is asked to assign a score from 1 to 10, 10 being the highest quality and 1 being the lowest. To make sure the evaluation is consistent, we use two models as the evaluator: Claude-3.7-sonnet and GPT-4.1-mini and compare the results.
% \footnote{The reason we did not chose the more common pair wise rating method is that we observed the transitivity does not hold for the model ranking: even when ground truth is give, we observe it is quite common that the evaluator judges response A is better than B, B is better than C, but C is better than A. Furthermore, this type of result is observed for different evaluator LLMs (Claude-3.7 sonnet, GPT-4o, GPT4.1-nano, GPT-4.1-mini). This should not have happened because the ground truth is provided. Therefore, we conclude that this result is not reliable.}.

\paragraph{Baselines.} We compare the five variants of \ours (four hypothetical response variants, one without hypothetical response) against the three modes of LightRAG~\citep{guo2024lightrag} (local, global, hybrid), three modes of GraphRAG~\citep{edge2024local} (local, global, drift), RQ-RAG~\citep{chan2024rq}, vanilla HyDE~\citep{gao2023precise} as well as vanilla RAG. In addition, we also give the hybrid mode of LightRAG the same four variants with the hypothetical response (thirteen in total excluding all the variants of our own methods). All methods use GPT-4o-mini \citep{hurst2024gpt} as the backbone LLM for preprocessing, inference, or both, and text-embedding-3-small \citep{text-embedding-3-small} as the embedding model \(f(\cdot)\).
% \paragraph{Evaluation.}  For our experiments ( \ref{sec:Experiments}) we grade model answers with GPT-4o, GPT-4.1-mini and Claude-3.7-Sonnet on a 1–10 Likert scale, following the rubric in Appendix~\ref{subsec:ResponsesEvaluationPrompt}.  Each sample is scored independently by all three models; we report the ranking based on the mean score across all three LLM judges in Table~\ref{tab:overall-avg-scores}. The individual judge's results can be found in Appendix~\ref{appendix:AdditionalResults}.

% \begin{equation}
% \textstyle
% \operatorname{Score}=
% \frac1J\sum_{j=1}^{J}\frac1N\sum_{i=1}^{N}s_{ij},
% \label{eq:eval}
% \end{equation}

\paragraph{Evaluation.}
For each query, we give an LLM judge the query, the ground truth and the responses of the variants of \ours and all baselines. Each answer is graded independently by \textit{J}~=~3 LLM judges
(Claude-3.7-Sonnet, GPT-4o, GPT-4.1-mini \citep{openai_gpt4.1_2025}) on a 1–10 Likert scale using the
rubric in Appendix \ref{subsec:ResponsesEvaluationPrompt}.  The overall
quality score for a system is the mean of those ratings across all
samples and judges, computed as

\begin{equation}
\operatorname{Score} \;=\;
\frac{1}{J}\sum_{j=1}^{J} \;
\frac{1}{N}\sum_{i=1}^{N} s_{ij},
\label{eq:eval}
\end{equation}

\noindent where $N$ is the number of question–answer pairs in the test set; $J$ is the number of LLM judges; and $s_{ij}\in\{1,\dots,10\}$ is the score assigned by judge $j$ to sample $i$.

We report the (\textit{mean score}) in Table \ref{tab:overall-avg-scores};
per-judge scores are provided in Appendix \ref{appendix:AdditionalResults}.

\paragraph{Result.} 
In the overall ranking, the top-3 modes are all hypothetical response variants of \ours, while \ours (vanilla) loses to GraphRAG local, GraphRAG drift and LightRAG hybrid mode. This corroborates that hypothetical responses play a major role in improving retrieval quality. In particular, the superior performance of the combined extraction variant, followed by the combined embedding variant, confirms that both incorporating the original query and employing the extraction step further enhance the retrieval effectiveness. In contrast, all hypothetical response variants of LightRAG hybrid mode performed poorly. We speculate that because \ours does not employ de-duplication, it synergizes particularly well with the hypothetical response, achieving top performance regardless of the specific variant used. Furthermore, the hypothetical response provides a richer context and an abundant number of entity and event candidates. This can be beneficial to the KG matching when there are event nodes to provide grounding to the entities to mitigate hallucination. Figure \ref{fig:heatmap} plots the category-average scores (averaged over the three evaluator LLMs).  In every category, the top-scoring model is again an \ours hypothetical response variant. For more details, refer to Appendix~\ref{appendix:AdditionalResults}.

\begin{table}[htbp]
  \centering
  \scriptsize
  \caption{Average query time per method (seconds)}
  \label{tab:avg_query_times}
  \small
  \begin{tabular}{lS}
    \toprule
    \textbf{Method} & {\textbf{Avg.\ Time}} \\
    \midrule
    GraphRAG drift                                      & 93.1547 \\
    GraphRAG global                                     & 26.2326 \\
    GraphRAG local                                      & 18.0141 \\
    RQ-RAG                                              & 9.9264  \\
    \ourss (comb. extraction)             & 8.7200  \\
    \ourss (hyp. embedding)        & 8.0850  \\
    \ourss (comb. embedding)              & 8.0544  \\
    \ourss (hyp. extraction)         & 8.0116  \\
    \ourss (vanilla)                                       & 7.5065  \\
    LightRAG hybrid (comb. extraction)          & 7.4924  \\
    LightRAG hybrid  (hyp. extraction)      & 7.1379  \\
    LightRAG hybrid (hypo. embedding)       & 6.7142  \\
    LightRAG hybrid (comb. embedding)           & 6.6951  \\
    LightRAG hybrid (vanilla)                                    & 6.1564  \\
    LightRAG local                                      & 5.2351  \\
    vanilla HyDE                                        & 2.9106  \\
    LightRAG global                                     & 1.9031  \\
    vanilla RAG                                         & 1.4224  \\
    \bottomrule
  \end{tabular}
\end{table}

\begin{table*}[!htbp]
  \tiny
  \setlength{\tabcolsep}{5pt}
  \caption{GPT-4.1-mini's verdicts (selected) of the responses to
    \emph{"Consider the dinner scene in which Holmes details his reasoning about the Lauriston Gardens mystery. How does Watson's narration highlight Holmes's eagerness to explain the logic step by step, and what rhetorical strategies (quoted or paraphrased) does Holmes use to underscore each clue's significance?"}}
  \label{tab:gpt41_verdict}
  \begin{tabularx}{\textwidth}{@{}l c X@{}}
    \toprule
    \textbf{Mode} & \textbf{Average Score} & \textbf{Reason} \\
    \midrule
    \zr \ourss (comb. extraction)   & 7.0000 & Very thorough and accurate, captures Watson's narration of Holmes's eagerness and detailed rhetorical strategies with direct quotes and logical progression, closely matching the ground truth.\\
    \zr LightRAG hybrid (comb. extraction) & 6.3333 & Captures Holmes's eagerness and rhetorical strategies well, including analogies, contrasts, and causal reasoning, with some direct quotes. However, it misses some of the specific step-by-step pacing and the theatrical, emphatic phrasing that the ground truth emphasizes, resulting in a somewhat less vivid depiction.\\
    
    \zr \ourss (vanilla)                              & 6.0000 & Detailed and faithful to the ground truth, includes Holmes's systematic approach, direct quotes, and rhetorical strategies, capturing both Watson's narration and Holmes's methodical explanation.\\

    \zr LightRAG hybrid                            & 6.0000 & Provides a thorough analysis of Watson's narration and Holmes's rhetorical strategies with examples and direct quotes, closely aligning with the ground truth's emphasis on step-by-step logic and Holmes's delight in explaining.\\
     GraphRAG drift                             & 6.0000 & Mentions Comprehensive and detailed, covers Watson's narration and Holmes's rhetorical strategies including analogies, rhetorical questions, and historical context, closely matching the ground truth.\\

         LightRAG local                             & 5.3333 & Describes Watson's narration and Holmes's rhetorical strategies like analogies and highlighting contradictions, but lacks specific direct quotes and detailed step-by-step logic as in the ground truth.\\

        vanilla RAG                                    & 4.3333 & Captures Holmes's eagerness and details several rhetorical strategies with some direct quotes and examples, but includes some inaccuracies and extraneous content not directly related to the Lauriston Gardens dinner scene.\\

    \zr vanilla HyDE                               & 3.6667 & Captures Holmes's eagerness and rhetorical strategies with some direct quotes and analogies, but less focused on the step-by-step logic and specific clues as in the ground truth.\\

    RQ-RAG                                         & 3.0000 & Very Very brief and vague; mentions Holmes's eagerness and rhetorical questions but lacks detail, examples, or direct quotes to support the answer.\\
  \end{tabularx}
\end{table*}

\paragraph{Preprocessing time cost.} Table~\ref{tab:pipeline-times} shows the preprocessing time of the novel \textit{The Phantom of the Opera} for three KG based RAG approaches used in our experiments.

% \begin{table}[ht]
%   \centering
%   \scriptsize
%   \caption{Preprocessing runtimes for the three KG RAG systems}
%   \label{tab:pipeline-times}

%   \resizebox{\columnwidth}{!}{%
%   \begin{tabular}{@{}l l r@{}}
%     \toprule
%     System & Stage & Time (s) \\ 
%     \midrule
%     LightRAG & Total pipeline & 101.1723 \\ 
%     \midrule
%     \multirow{4}{*}{Entity--Event}
%       & Entity subgraph construction   & 108.3549 \\ 
%       & Event subgraph construction    & 103.6091 \\ 
%       & Bipartite-mapping       &   0.5076 \\
%       & \textbf{Total pipeline}        & 108.8625\\
%     \midrule
%     GraphRAG & Total pipeline & 189.0648 \\ 
%     \bottomrule
%   \end{tabular}}%
% \end{table}

 Compared to LightRAG, although \ours needs to construct two subgraphs \(
  (\mathcal{G}_{\text{ent}},\mathcal{G}_{\text{evt}})
\) as opposed to just event KG, and form \(
\mathcal{B}
\) (Bipartite mapping), the two subgraphs' constructions are independent and can be carried out in parallel. Forming \(
\mathcal{B}\) is sequential after the subgraphs' construction, but for \((\mathcal{G}_{\text{ent}},\mathcal{G}_{\text{evt}})
\) of reasonable size such as in this example, its cost is negligible in practice. As a result, \ours has preprocessing time comparable to LightRAG (subject to API query traffic fluctuation), whereas GraphRAG is noticeably slower.

\paragraph{Query time cost.}
% \begin{figure*}[!ht]
%     \centering
%     \includegraphics[width=0.8\linewidth]{avg_query_times.png}
%     \caption{
%         Average query time of each mode.
%     }
%     \label{fig:avg_query_times}
% \end{figure*}

Table~\ref{tab:avg_query_times} shows the average query time for each mode. GraphRAG drift takes the longest time. \ours (vanilla) takes marginally longer time compared to LightRAG hybrid mode, likely due to the time taken for searching the additional events on KG. The hypothetical response variants all took slightly longer time than their respective base form due to the additional step of generating the hypothetical response. In particular, the combined extraction variant took the longest time due to the extraction process. Nevertheless, the overall query time difference for LightRAG and \ours is insignificant with or without hypothetical response, and much shorter compared to all GraphRAG modes.

\paragraph{Token Cost Analysis.} When KG construction phase, since event extraction is done separately from entity extraction, \ours has two times the number of API calls of LightRAG with same max token count $C_{\text{extract}}$, which doubles its worst-case token cost during preprocessing. On the other hand, as the author of LightRAG pointed out, GraphRAG is still has much higher token consumption when comes to KG construction due to massive, repeatedly generated community reports \citep{guo2024lightrag} (refer to Table~\ref{tab:token-usage} for the cost on preprocessing \textit{The Phantom of the Opera}). During retrieval, to ensure a fair comparison, we set the max token $C_{\text{output}}$ for the retrieved chunks to be the same for every KG-based RAG mode; therefore, they all have the same toke cost. For every query, the hypothetical response step contributes to exactly one additional API call with the same $C_{\text{output}}$ allowed per API call, which doubles the output token count in the worst case.

\paragraph{Case study.}
In Table~\ref{tab:gpt41_verdict}, we showcase the responses selected modes used for evaluation for a particular query from \textit{The Hound of the Baskervilles} and the verdicts given by GPT-4.1-mini. The complete verdicts of all three LLM judges can be found in Appendix~\ref{appendix:verdicts}. Overall, pairing with hypothetical response, \ours is able to retrieve contextually relevant chunks—rather than surface-level matches—yielding more thorough and accurate answers. The details on the extraction and retrieval of \ours (comb. extraction) mode can be found in Appendix~\ref{appendix:case_study}. Due to the space constraints, complete responses of other modes are provided in the supplementary materials.

\section{Conclusion}
\ours keeps every entity mention separate and anchors it to the exact events in which it appears, restoring the temporal and causal context that vanilla RAG and deduplicated KG variants fail to capture.  On the new ChronoQA benchmark—designed specifically to test narrative, causal, and character-consistency reasoning—it delivers the best overall and category-specific scores while matching LightRAG's preprocessing cost and latency.  Because it builds on off-the-shelf extraction prompts and vector stores, requires no model fine-tuning, and pairs naturally with hypothetical response, \ours offers a drop-in upgrade for more faithful retrieval and a foundation for future work on RAG for complex question-answering tasks.

\section*{Limitations}
The proposed \ours\ framework focuses on improving the response  quality of the retrieval-augmented generation on documents that possess inherent temporal or causal structures—for instance, novels, short stories, scripts, and other narrative-style texts in which characters evolve and events unfold in sequence. Outside of this scope, where information lacks such structures, \ours\ may not offer noticeable benefits compared to methods specifically designed for those contexts.

% \clearpage
\bibliography{custom}

\clearpage

\appendix
\onecolumn

% ---------- Appendix A. chronoQA Statistics ----------
\section{More ChronoQA Details}
\label{appendix:Statistics}
\addcontentsline{toc}{section}{Appendix A. chronoQA Statistics}

\paragraph{Format and release.} Each record is a JSON line with fields
\texttt{\{ "story\_id", "story\_title", "question\_id", "category", "question", "ground\_truth", "passages" \}},  
where \texttt{passages} has subfields \texttt{\{ "start\_sentence", "end\_sentence", "start\_byte", "end\_byte", "excerpt"\}} which pinpoint the relevant passage(s) from the story that provide the supporting evidence of the ground truth.  The generation prompt can be found in Appendix \ref{subsec:DataGenerationPrompts}.

\section{Preprocessing Token Cost}
% Token-usage table
\begin{table}[ht]
  \centering
  \caption{Token usage statistics for the three KG-RAG systems}
  \label{tab:token-usage}
  \resizebox{0.4\linewidth}{!}{%
    \begin{tabular}{@{}l
                S[table-format=7.0,
                  round-mode=places,
                  round-precision=0]  % input
                S[table-format=7.0,
                  round-mode=places,
                  round-precision=0]  % output
                S[table-format=8.0,
                  round-mode=places,
                  round-precision=0]@{}}  % total
  \toprule
  System   & {Input tokens} & {Output tokens} & {Total tokens} \\
  \midrule
  LightRAG & 343437  & 112840  & 456277 \\
  \ourss   & 598822  & 208797  & 807619 \\
  GraphRAG & 880804  & 440401  & 1321205 \\
  \bottomrule
\end{tabular}
  }
\end{table}

\section{Hyperparameters}

We implement \ours within LightRAG's existing codebase. For all experiments, we standardize the chunk size, max token for backbone LLM generation, and chunks appended to be the same as the default setting as LightRAG. We experimented on changing their values but observed no noticeable benefit.

\section{Prompts}
\label{sec:Prompts}

\subsection{Data Generation Prompt}
\label{subsec:DataGenerationPrompts}
\begin{DocChunk}
I want to test a few rag systems on their reasoning capabilities and the capability to pick up nuanced details. Use the document shown below, design queries for it, also provide the ground truth for each query. In addition, it should focus on testing the RAG system's causal consistency. for example, 1. it can ask the RAG system how would a certain character behavior in a given context/even from the document, and see if the answer is consistent with the character's personality/traits at that specific point in time (character can experience development so their response to the same thing can very). 2. it can test if the RAG system confuse the event that has not happened but in the document with the query (the future events that has not happen should not be account for in the answer). Give the queries and ground truth in json format.the document is here:

DOCUMENT

Focus on Causal Consistency: Apart from character behavior over time and future event confusion, other causal consistency tests (e.g., testing how well it understands cause-and-effect relationships in the plot) can also be included. Query Difficulty: query should involve complex, nuanced reasoning/understanding of the document provided.
Format example:
{
  "queries": [
    {
      "query": "During Harry's first night at Hogwarts, the Gryffindor students climb the moving staircases. How does the text describe the corridors and stairways' magical behavior, and what is the immediate impact on Harry's sense of direction?",
      "ground_truth": "The staircases sometimes change direction, doors can vanish or move, and some require a password or a specific tickle of a doorknob to open. This constant shifting confuses new students like Harry, making it easy to get lost early on.",
      "start_sentence": "The staircases at Hogwarts are famous for moving unexpectedly, often depositing unwary students on entirely different floors than intended.",
      "end_sentence": "This enchantment leaves first-years such as Harry feeling hopelessly lost during their first nights in the castle.",
      "type": "Causal Consistency"
    }
  ]
}

Rules for the `start_sentence` and `end_sentence` strings:
• They must be *identical substrings* of the document (case‑sensitive, byte‑for‑byte).
• Preserve every original character: spaces, line‑breaks, hyphens, quotation marks, etc.
• If the document contains line breaks, represent them in JSON as the two-character sequence.
• Do not add, delete, or normalise any characters—copy‑paste only.
• The passage between the two sentences must support the ground‑truth answer you give.

It is also important to note that do not explicitly disclose the title/chapter/section number from which the context of the question is used. Simply described the related event and the characters invoved to make the context clear.
\end{DocChunk}

\subsection{Responses Evaluation Prompt}
\label{subsec:ResponsesEvaluationPrompt}
\begin{DocChunk}
You are an expert evaluator of retrieval‑augmented generation (RAG) answers.

Scoring rubric (10‑point scale):
10 – Matches ground truth exactly or with faithful paraphrase.
7  – Mostly correct; minor omissions or wording differences.
5  – Partially correct; major missing points or inaccuracies.
3  – Mostly incorrect; small overlap.
1  – Off‑topic or hallucinated.

Return **only** a valid JSON array, no markdown fences, in this exact shape:
[
  {"mode": "mode\_name", "reason": "short rationale", "score": 9},
  ...
]

If you cannot produce the JSON array, return an object like:
{"error": "description"}.
\end{DocChunk}

\subsection{Entity, Event Extraction Prompt}
\label{subsec:ExtractionPrompt}
\begin{DocChunk}
---Role---

You are a helpful assistant tasked with identifying entities and events in the user's query.

---Goal---

Given the query, list both entities and events. Entities are people, places, organizations, or objects mentioned in the query, while events are actions, occurrences, or happenings that take place.

---Instructions---

- Output the entities and events in JSON format.
- The JSON should have two keys:
  - "entities" for people, places, organizations, or objects.
  - "events" for actions, occurrences, or happenings.

######################
-Examples-
######################
Example 1:

Query: "How did Napoleon's invasion of Russia affect his empire's strength?"
################
Output:
{{
  "entities": ["Napoleon", "Russia", "Napoleon's empire"],
  "events": ["invasion of Russia", "empire's decline"]
}}
#############################
Example 2:

Query: "What role did MIT scientists play in the Manhattan Project?"
################
Output:
{{
  "entities": ["MIT", "MIT scientists", "Manhattan Project"],
  "events": ["scientific research", "atomic bomb development"]
}}
#############################
Example 3:

Query: "How did the Industrial Revolution change London's population?"
################
Output:
{{
  "entities": ["London", "London's population", "Industrial Revolution"],
  "events": ["population growth", "urbanization", "industrial development"]
}}
#############################
-Real Data-
######################
Query: {query}
######################
Output:

"""

PROMPTS["naive_rag_response"] = """---Role---

You are a helpful assistant responding to questions about documents provided.

---Goal---

Generate a response of the target length and format that responds to the user's question, summarizing all information in the input data tables appropriate for the response length and format, and incorporating any relevant general knowledge.
If you don't know the answer, just say so. Do not make anything up.
Do not include information where the supporting evidence for it is not provided.

---Target response length and format---

{response_type}

---Documents---

{content_data}

Add sections and commentary to the response as appropriate for the length and format. Style the response in markdown.
\end{DocChunk}

\section{Entity-Event KG Visualization}
\begin{figure*}[!ht]
    \centering
    \includegraphics[width=1.0\linewidth]{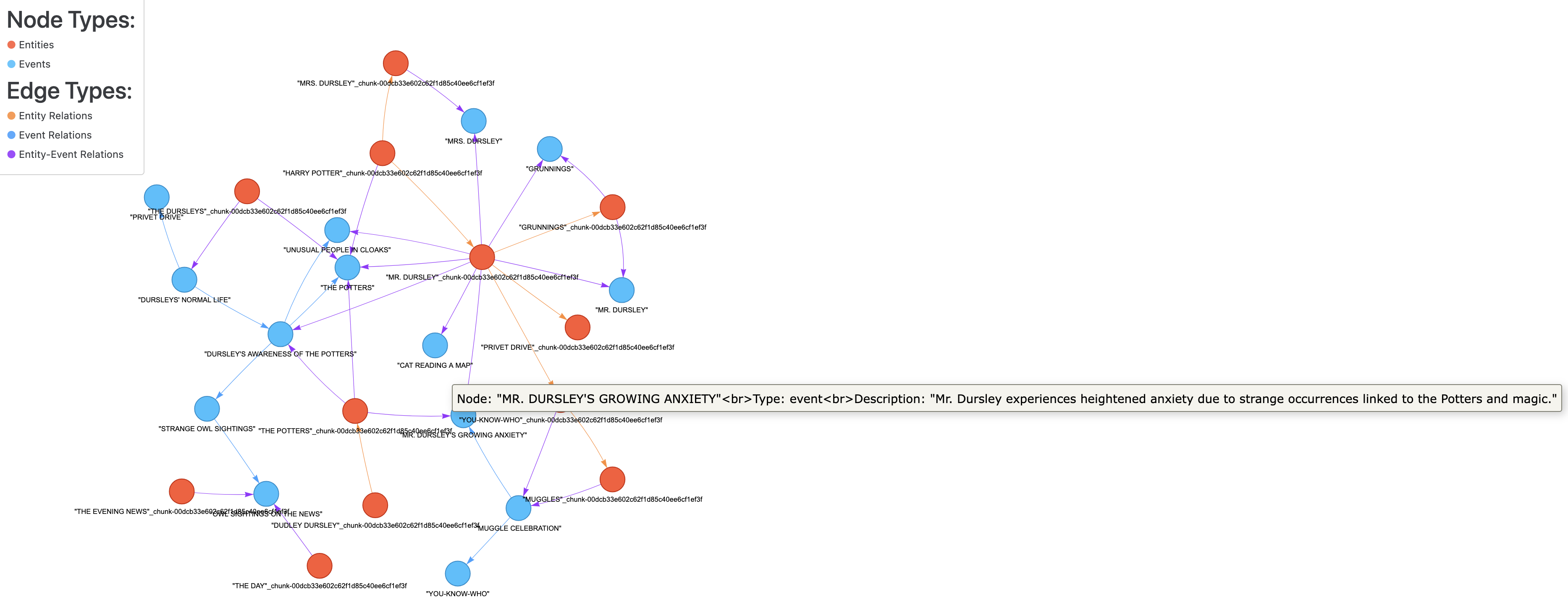}
    \caption{
        A snapshot of a part of the Entity-Event KG of \textit{Harry Potter and the Sorcerer's Stone}.
    }
    \label{fig:e2-KG}
\end{figure*}

\section{More Details on Case Study on Mode: \ours (comb. extraction)}
\label{appendix:case_study}

\subsection{Hypothetical Response}
\begin{DocChunk}
In the dinner scene from "The Hound of the Baskervilles," Watson's narration captures Holmes\'s fervent enthusiasm for elucidating his reasoning behind the Lauriston Gardens mystery. As they sit at the table, Watson observes Holmes's animated demeanor, noting how "his eyes shone with a keen light" and how he "leaned forward with an eagerness that betrayed his excitement." This vivid description not only illustrates Holmes\'s passion for deduction but also emphasizes his desire to engage Watson in the unraveling of the case, making him feel like a partner in the detective\'s intellectual adventure.\n\nHolmes employs various rhetorical strategies to accentuate the significance of each clue as he meticulously lays out his thought process. For instance, he often uses analogies to simplify complex ideas, comparing the clues to pieces of a puzzle that, when assembled correctly, reveal the bigger picture. He states, "The world is full of obvious things which nobody by any chance ever observes," underscoring how critical attention to detail is in investigative work. This rhetorical device serves to remind Watson-and the reader-of the importance of keen observation in deciphering the mystery.\n\nMoreover, Holmes frequently employs the technique of building suspense through his methodical breakdown of the evidence. He carefully dissects each piece of information, pausing for effect before revealing its implications. By saying, "And yet, it is the most obvious thing in the world," he not only highlights the clarity of his deductions but also invites the audience to share in the thrill of discovery. This step-by-step approach not only showcases Holmes\'s logical prowess but also enhances Watson's narrative by creating a sense of shared enlightenment, as they both traverse the landscape of clues together.\n\nUltimately, Watson\'s narration, rich in detail and observation, amplifies Holmes's eagerness to teach and elucidate, while Holmes's rhetorical strategies serve to elevate the weight of each clue, transforming the dinner scene into a dynamic exploration of deduction and insight.
\end{DocChunk}

\subsection{Extracted Data}

Extracted entity keywords: 
\begin{DocChunk}
['Watson', 'dinner scene', 'The Hound of the Baskervilles', 'Lauriston Gardens', 'Holmes']
\end{DocChunk}

Extracted event keywords: 
\begin{DocChunk}
["Holmes's enthusiasm", 'dinner scene', 'discussion of the Lauriston Gardens mystery', 'breakdown of evidence', 'building suspense', "underscoring clues' significance", "narration of Holmes's reasoning", 'detailing reasoning', 'highlighting eagerness', 'explaining logic', 'using rhetorical strategies', 'shared enlightenment', 'deduction process']
\end{DocChunk}

Retrieved nodes data: 
\begin{DocChunk}
[\{'type': '"PERSON"', 'description': '"Holmes is a character portrayed as a detective who engages in solving a mystery surr...and expresses confidence in his theories."', 'source\_id': 'chunk-e16ab5c062904b8574746080e70db172', 'node\_type': 'entity', 'display\_color': '#FF5733', 'entity\_name': '"HOLMES"\_chunk-e16ab5c062904b8574746080e70db172', 'rank': 9\}, ...\{'type': '"EVENT"', 'description': '"Holmes elaborates on his strategy to catch the assassin while navigating the complexities of the situation."', 'source\_id': 'chunk-fbff6479d4cfc0843851b4ff06a86927', 'node\_type': 'event', 'display\_color': '#33C1FF', 'entity\_name': '"HOLMES\'S STRATEGY DISCUSSION"', 'rank': 2\}...]
\end{DocChunk}
Retrieved edge data: 
\begin{DocChunk}
[\{'src\_tgt': (...), 'rank': 37, 'weight': 8.0, 'description': '"I engages in dialogue with Holmes, seeking understanding of the detective\'s methodologies and deductions."', 'keywords': '"narration, learning"', 'source\_id': 'chunk-9d367c150b511e67b26c6a5701e7fcb9', 'edge\_type': 'entity\_relation', 'display\_color': '#FF8C33'\}, \{'src\_tgt': (...), 'rank': 25, 'edge\_type': 'entity\_event\_relation', 'description': 'Entity \'"Holmes is a skilled detective known for his sharp powers of observation and ...s wheels as a crucial piece of evidence."\'', 'weight': 1.0, 'keywords': 'ENTITY\_EVENT\_RELATION', 'source\_id': 'chunk-9d367c150b511e67b26c6a5701e7fcb9<SEP>chunk-9d367c150b511e67b26c6a5701e7fcb9', 'display\_color': '#9933FF'\},...]
\end{DocChunk}

Retrieved context data (relevant chunks): 
\begin{DocChunk}
\n-----Sources-----\n```csv\nid,content\r\n0,"and that the depravity of the victim was no condonement in the eyes of the law.\n\nThe more I thought of it the more extraordinary did my companion's hypothesis, that the man had been poisoned, appear. I remembered how he had sniffed his lips, and had no doubt that he had detected something which had given rise to the idea. Then, again, if not poison, what had caused the man's death, since there was neither wound nor marks of strangulation? But, on the other hand, whose blood was that which lay so thickly upon the floor? There were no signs of a struggle, nor had the victim any weapon with which he might have wounded an antagonist. As long as all these questions were unsolved, I felt that sleep would be no easy matter, either for Holmes or myself. His quiet self-confident manner convinced me that he had already formed a theory which explained all the facts, though what it was I could not for an instant conjecture. ...
\end{DocChunk}
Final response:
\begin{DocChunk}
In the dinner scene where Sherlock Holmes discusses the Lauriston Gardens mystery with Dr. Watson, Watson \u2019s narration reveals Holmes's enthusiastic and methodical approach to detective work. This moment underscores not only Holmes\u2019s intellectual prowess but also his eagerness to engage Watson in the reasoning process that leads to his deductions. Watson\u2019s observations during this discussion reflect his genuine admiration for Holmes's analytical mind, which illuminates the depth of the detective\u2019s insight and reasoning.\n\n### Watson\u2019s Narrative Perspective\n\nWatson, as the narrator, captures the intensity of Holmes\u2019s engagement with the case. He notes how Holmes displays a \"quiet self-confident manner,\" suggesting that Holmes is not only certain of his deductions but also eager to articulate the thought process that led him there. For instance, Watson recounts moments where Holmes transitions on topics with excitement, indicating how much he relishes sharing the intricacies of his method \u2014 \u201cThe plot thickens,\u201d he exclaims at key junctures, revealing his thrill at unraveling the mystery. This narrative style enhances Holmes's character as not just a brilliant detective but also a passionate educator, inviting Watson\u2014and, by extension, the reader\u2014into his deductive world.
\n\n### Rhetorical Strategies Used by Holmes\n\nHolmes employs several rhetorical strategies to underscore the significance of each clue in his explanation. These strategies include:\n\n1. **Logical Sequencing**: Holmes emphasizes the importance of reasoning backwards and forwards, which he articulately illustrates when he examines the physical evidence. For example, he reasons from the cab\u2019s ruts and horse\u2019s hoof prints left on the wet soil to deduce that two men must have visited the house. Here, he insists that \"it is a capital mistake to theorize before you have all the evidence,\" highlighting his disciplined approach to deduction that relies on empirical observations rather than mere speculation.\n\n2. **Detailed Observation**: Holmes's acute observations serve as pivotal points in his reasoning. He recognizes that the \"very first thing\" he observed was the impressions left by the cab wheels. By meticulously detailing this observation, he conveys to Watson\u2014and the audience\u2014the fundamental role of minute details in solving a case: \"There is no branch of detective science which is so important and so much neglected as the art of tracing footsteps.\"\n\n3. **Analytical Clarity**: As he breaks down how he calculated the height and nature of the suspects based on their footprints and style of dress, Holmes uses clear, methodical language. He moves from discussing the physical evidence to the psychological implications, illustrating how the spatial and temporal context plays a critical role in understanding the crime. For example, he reflects on the footprints and concludes, \u201cIt is simple enough\u2026 I could read all that in the dust.\u201d \n\n4. **Connecting to Broader Concepts**: Holmes often draws on broader concepts of criminality and human behavior, thereby situating his deductions within a larger framework of detective work. He references historical cases and theories that bolster his arguments, explaining his reasoning not just as a technical endeavor, but as a complex interplay of human motivations and actions.\n\n### Conclusion\n\nHolmes's eagerness to share his deductive reasoning exemplifies the detective's character as someone deeply engaged in the pursuit of knowledge and understanding. Through Watson\u2019s narration, readers see an interplay of admiration, curiosity, and insight as Holmes articulates the significance of each clue. The combination of logical sequencing, detailed observation, analytical clarity, and broader contextual reasoning allows Holmes to construct a compelling narrative that not only uncovers the truth but also elevates the art of detection itself. This dinner scene, rich with multi-layered reasoning, highlights the intricacies of Holmes's mind, drawing readers closer to the fascinating world of deduction.
\end{DocChunk}

\clearpage
\subsection{Complete Verdicts from All Three LLM Judges}
\label{appendix:verdicts}

\begin{table*}[!htbp]
  \footnotesize
  \setlength{\tabcolsep}{5pt}
  \caption{GPT-4.1-mini's verdicts (part 1) of the responses to
    \emph{"Consider the dinner scene in which Holmes details his reasoning about the Lauriston Gardens mystery. How does Watson's narration highlight Holmes's eagerness to explain the logic step by step, and what rhetorical strategies (quoted or paraphrased) does Holmes use to underscore each clue's significance?"}}
  \label{tab:gpt41_verdict_complete}
  \begin{tabularx}{\textwidth}{@{}l c X@{}}
    \toprule
    \textbf{Mode} & \textbf{Average Score} & \textbf{Reason} \\
    \midrule
    \zr \ourss (comb. extraction)   & 7.0000 & Very thorough and accurate, captures Watson's narration of Holmes's eagerness and detailed rhetorical strategies with direct quotes and logical progression, closely matching the ground truth.\\
    \ourss (comb. embedding)        & 7.0000 & Comprehensive and well-aligned with the ground truth, includes detailed analysis of Watson's narration, Holmes's eagerness, and rhetorical strategies with direct quotes.\\
    \zr    \ourss (hyp. embedding)    & 7.0000 & Detailed and faithful, includes Holmes's eagerness, rhetorical strategies, direct quotes, and Watson's narration, effectively capturing the essence of the ground truth.\\
    \zr LightRAG hybrid (hyp. extraction) & 6.6667 & Highly detailed and faithful to the ground truth, includes Holmes's systematic reasoning, direct quotes, rhetorical questions, and Watson's narration highlighting Holmes's eagerness.\\
    \zr LightRAG hybrid (comb. extraction) & 6.3333 & Captures Holmes's eagerness and rhetorical strategies well, including analogies, contrasts, and causal reasoning, with some direct quotes. However, it misses some of the specific step-by-step pacing and the theatrical, emphatic phrasing that the ground truth emphasizes, resulting in a somewhat less vivid depiction.\\
     \ourss (hyp. extraction) & 6.3333 & Provides a detailed and faithful account of Watson's narration and Holmes's eagerness, including direct quotes and rhetorical strategies such as deductive reasoning, metaphor, and analogy. However, it lacks some of the exact emphatic openings, instructive questions, and dismissive flourishes explicitly cited in the ground truth, making it slightly less precise.\\
     
     \zr LightRAG hybrid (comb. embedding)      & 6.3333 & Providing a solid overview of Holmes's eagerness and rhetorical strategies with direct quotes and logical reasoning. However, it does not fully capture the breathless pacing and specific emphatic or dismissive phrases that characterize the ground truth.\\
    
    LightRAG hybrid (hyp. embedding)  & 6.3333 & Nearly identical to hybrid\_combined\_extraction in content and style, with good coverage of Watson's narration and Holmes's rhetorical strategies but lacking the full range of direct quotes and the theatrical emphasis present in the ground truth.\\
    
    \zr \ourss (vanilla)                                   & 6.0000 & Detailed and faithful to the ground truth, includes Holmes's systematic approach, direct quotes, and rhetorical strategies, capturing both Watson's narration and Holmes's methodical explanation.\\
     GraphRAG global                            & 6.0000 & Addresses Strong answer highlighting Holmes's eagerness and rhetorical strategies like analogies and emphatic language, with direct quotes and good alignment to the ground truth.\\

    \zr LightRAG hybrid                            & 6.0000 & Provides a thorough analysis of Watson's narration and Holmes's rhetorical strategies with examples and direct quotes, closely aligning with the ground truth's emphasis on step-by-step logic and Holmes's delight in explaining.\\
     GraphRAG drift                             & 6.0000 & Mentions Comprehensive and detailed, covers Watson's narration and Holmes's rhetorical strategies including analogies, rhetorical questions, and historical context, closely matching the ground truth.\\
     \zr  GraphRAG local                                 & 5.6667 & Good summary of Watson's narration and Holmes's rhetorical strategies including logical reasoning and analogies, with some direct quotes, but less detailed than the ground truth.\\
      \zr LightRAG global                                & 5.6667 & Effectively highlights Holmes's eagerness and rhetorical strategies such as analogies and rhetorical questions with some direct quotes, closely matching the ground truth's focus on Holmes's methodical explanation.\\
      
  \end{tabularx}
\end{table*}

% ---------- second half ----------
\begin{table*}[!htbp]\ContinuedFloat   % <— keeps the same number
  \footnotesize
  \setlength{\tabcolsep}{5pt}
  \caption{GPT-4.1-minis verdicts (continued)}
  \begin{tabularx}{\textwidth}{@{}l c X@{}}
    \toprule
    \textbf{Mode} & \textbf{Average Score} & \textbf{Reason}\\
    \midrule       
    
         LightRAG local                             & 5.3333 & Describes Watson's narration and Holmes's rhetorical strategies like analogies and highlighting contradictions, but lacks specific direct quotes and detailed step-by-step logic as in the ground truth.\\
  
        vanilla RAG                                    & 4.3333 & Captures Holmes's eagerness and details several rhetorical strategies with some direct quotes and examples, but includes some inaccuracies and extraneous content not directly related to the Lauriston Gardens dinner scene.\\

    \zr vanilla HyDE                               & 3.6667 & Captures Holmes's eagerness and rhetorical strategies with some direct quotes and analogies, but less focused on the step-by-step logic and specific clues as in the ground truth.\\

    RQ-RAG                                         & 3.0000 & Very Very brief and vague; mentions Holmes's eagerness and rhetorical questions but lacks detail, examples, or direct quotes to support the answer.\\
    \bottomrule
  \end{tabularx}
\end{table*}

\begin{table*}[!htbp]
  \footnotesize
  \setlength{\tabcolsep}{5pt}
  \caption{Claude-3.7-Sonnet's verdicts (part 1) of the responses to
    \emph{"Consider the dinner scene in which Holmes details his reasoning about the Lauriston Gardens mystery. How does Watson's narration highlight Holmes's eagerness to explain the logic step by step, and what rhetorical strategies (quoted or paraphrased) does Holmes use to underscore each clue's significance?"}}
  \label{tab:claude_verdict}
  \begin{tabularx}{\textwidth}{@{}l c X@{}}
    \toprule
    \textbf{Mode} & \textbf{Average Score} & \textbf{Reason} \\
    \midrule
    \zr \ourss (comb. extraction)   & 7.0000 & Mentions footprints and cab marks, and discusses Holmes's eagerness to explain his reasoning. However, it doesn't capture the specific rhetorical pattern of short declarations followed by challenges to Watson.\\
    \ourss (comb. embedding)        & 7.0000 & Discusses Holmes's eagerness and mentions footprints and stride length, but doesn't fully capture the rhetorical pattern of short declarations followed by challenges to Watson described in the ground truth.\\
    \zr    \ourss (hyp. embedding)    & 7.0000 & Mentions cab marks and footprints, and discusses Holmes's eagerness to explain his reasoning. However, it doesn't capture the specific rhetorical pattern of short declarations followed by challenges to Watson.\\
    \zr LightRAG hybrid (hyp. extraction) & 6.6667 & Mentions key elements like cab wheel ruts and stride length, and discusses Holmes's eagerness to explain his reasoning. Includes some specific rhetorical strategies but doesn't fully capture the rapid, orderly exposition and the pattern of declarations followed by challenges to Watson described in the ground truth.\\
    \zr LightRAG hybrid (comb. extraction) & 6.3333 & Accurately mentions cab tracks and Holmes's eagerness to explain his reasoning step by step. References some rhetorical strategies like analogies and contrasts, but doesn't fully capture all the specific elements (fingernails, puddle width) or the rhetorical pattern of emphatic openings and dismissive flourishes in the ground truth.\\
     \ourss (hyp. extraction) & 6.3333 & Mentions cab marks and footprints but doesn't specifically address Holmes's eagerness to explain step by step or his rhetorical pattern of making declarations followed by challenging Watson.\\
     
     \zr LightRAG hybrid (comb. embedding)      & 6.3333 & Mentions cab tracks and Holmes's eagerness to explain his reasoning. Includes some rhetorical strategies but doesn't fully capture all the specific elements or the rhetorical pattern described in the ground truth.\\
    
    LightRAG hybrid (hyp. embedding)  & 6.3333 & Mentions cab tracks and Holmes's eagerness to explain his reasoning. Includes some rhetorical strategies but doesn't fully capture all the specific elements or the rhetorical pattern described in the ground truth.\\
    
    \zr \ourss (vanilla)                                   & 6.0000 & Mentions some key elements like footprints and cab marks but doesn't specifically address Holmes's eagerness to explain step by step or his rhetorical pattern of making declarations followed by challenging Watson.\\
     GraphRAG global                            & 6.0000 & Addresses Holmes's eagerness to explain his reasoning but lacks specific references to footprints, stride lengths, and cab marks. The rhetorical strategies mentioned are not specific to the text.\\

    \zr LightRAG hybrid                            & 6.0000 & Provides a detailed analysis of Holmes's eagerness and rhetorical strategies but doesn't specifically mention the footprints, stride lengths, and cab marks that are central to the ground truth. The rhetorical strategies are more generalized than specific.\\
     GraphRAG drift                             & 6.0000 & Mentions Holmes's eagerness and some rhetorical strategies but lacks specific references to footprints, stride lengths, and cab marks. The analysis is somewhat generic rather than text-specific.\\
      
    % ---- rows 1-k ----
  \end{tabularx}
\end{table*}

% ---------- second half ----------
\begin{table*}[!htbp]\ContinuedFloat   % <— keeps the same number
  \footnotesize
  \setlength{\tabcolsep}{5pt}
  \caption{Claude-3.7-Sonnet's verdicts (continued)}
  \begin{tabularx}{\textwidth}{@{}l c X@{}}
    \toprule
    \textbf{Mode} & \textbf{Average Score} & \textbf{Reason}\\
    \midrule       
    \zr  GraphRAG local                                 & 5.6667 & Discusses Holmes's eagerness and some rhetorical strategies but lacks specific references to the footprints, stride lengths, and cab marks mentioned in the ground truth. The analysis is somewhat generic.\\
         LightRAG local                             & 5.3333 & Discusses Holmes's eagerness and methodical reasoning but lacks specific references to footprints, stride lengths, and cab marks mentioned in the ground truth. The rhetorical strategies are generalized rather than specific to the text.\\
    \zr LightRAG global                                & 5.6667 & Addresses Holmes's eagerness to explain his logic but lacks specific references to the footprints, stride lengths, and cab marks. Mentions rhetorical questions but doesn't capture Holmes's pattern of short declarations followed by challenges to Watson.\\
        vanilla RAG                                    & 4.3333 & Contains some relevant elements about Holmes's analytical reasoning but misquotes Holmes and includes fabricated dialogue. The analysis lacks specific references to Watson's narration of Holmes's eagerness and the step-by-step logic mentioned in the ground truth.\\

    \zr vanilla HyDE                               & 3.6667 & Discusses a different story ('The Hound of the Baskervilles') and doesn't mention the specific elements from the ground truth like footprints, stride lengths, and cab marks.\\

    RQ-RAG                                         & 3.0000 & Very brief and generic. While it mentions Holmes's eagerness and rhetorical strategies, it lacks specific details about footprints, stride lengths, and cab marks. The answer is too vague and underdeveloped.\\
    \bottomrule
  \end{tabularx}
\end{table*}

\begin{table*}[!htbp]
  \footnotesize
  \setlength{\tabcolsep}{5pt}
  \caption{GPT-4o's verdicts (part 1) of the responses to
    \emph{"Consider the dinner scene in which Holmes details his reasoning about the Lauriston Gardens mystery. How does Watson's narration highlight Holmes's eagerness to explain the logic step by step, and what rhetorical strategies (quoted or paraphrased) does Holmes use to underscore each clue's significance?"}}
  \label{tab:gpt4o_verdict}
  \begin{tabularx}{\textwidth}{@{}l c X@{}}
    \toprule
    \textbf{Mode} & \textbf{Average Score} & \textbf{Reason} \\
    \midrule
    \zr \ourss (comb. extraction)   & 7.0000 & Partially correct; discusses Holmes's eagerness and some rhetorical strategies but lacks specific examples.\\
    \ourss (comb. embedding)        & 7.0000 & Partially correct; discusses Holmes's eagerness and some rhetorical strategies but lacks specific examples.\\
    \zr    \ourss (hyp. embedding)    & 7.0000 & Partially correct; discusses Holmes's eagerness and some rhetorical strategies but lacks specific examples.\\
    \zr LightRAG hybrid (hyp. extraction) & 6.6667 & Partially correct; discusses Holmes's eagerness and some rhetorical strategies but lacks specific examples.\\
    \zr LightRAG hybrid (comb. extraction) & 6.3333 & Partially correct; discusses Holmes's eagerness and some rhetorical strategies but lacks specific examples.\\
     \ourss (hyp. extraction) & 6.3333 & Partially correct; discusses Holmes's eagerness and some rhetorical strategies but lacks specific examples.\\
     
     \zr LightRAG hybrid (comb. embedding)      & 6.3333 & Partially correct; discusses Holmes's eagerness and some rhetorical strategies but lacks specific examples.\\
    
    LightRAG hybrid (hyp. embedding)  & 6.3333 & Partially correct; discusses Holmes's eagerness and some rhetorical strategies but lacks specific examples.\\
    
    \zr \ourss (vanilla)                                   & 6.0000 & Partially correct; discusses Holmes's enthusiasm and some rhetorical strategies but lacks specific examples.\\
     GraphRAG global                            & 6.0000 & Addresses Partially correct; discusses Holmes's eagerness and some rhetorical strategies but lacks specific examples.\\

    \zr LightRAG hybrid                            & 6.0000 & Partially correct; mentions Holmes's eagerness and some rhetorical strategies but lacks specific examples.\\
     GraphRAG drift                             & 6.0000 & Mentions Partially correct; discusses Holmes's eagerness and some rhetorical strategies but lacks specific examples.\\
      \zr  GraphRAG local                                 & 5.6667 & Partially correct; discusses Holmes's eagerness and some rhetorical strategies but lacks specific examples.\\
         LightRAG local                             & 5.3333 & Partially correct; mentions Holmes's enthusiasm and some rhetorical strategies but lacks specific examples.\\
 
    \zr LightRAG global                                & 5.6667 & Partially correct; discusses Holmes's eagerness and rhetorical strategies but lacks specific examples from the scene.\\
        vanilla RAG                                    & 4.3333 & Mostly incorrect; lacks specific details about Holmes's rhetorical strategies and Watson's narration.\\

    \zr vanilla HyDE                               & 3.6667 & Off-topic; discusses 'The Hound of the Baskervilles' instead of Lauriston Gardens.\\

    RQ-RAG                                         & 3.0000 & Very Mostly incorrect; brief mention of rhetorical strategies without specific examples or context.\\
    \bottomrule
  \end{tabularx}
\end{table*}

\newpage

\section{Additional Results}
\label{appendix:AdditionalResults}

Table~\ref{tab:avg-scores-claude} to Table~\ref{tab:avg-scores-gpt4mini} show the ranking results of each individual LLM evaluator. It can be observed each mode's ranking remains relatively stable. On the other hand, different evaluators have different standards, as the scores given by Claude-3-7-sonnet are generally low while the scores given by GPT-4.1-mini are generally high across the board. Table~\ref{tab:causal-consistency} to Table~\ref{tab:thematic} show the ranking results for the questions based on category, averaged across all three evaluator LLMs.

%===================== Claude-3-7 judge =====================
\begin{table}[htbp]
\centering
\small
\caption{Average scores for \texttt{Claude-3-7-sonnet}}
\label{tab:avg-scores-claude}

\rowcolors{3}{gray!10}{white}
\begin{tabular}{
  @{}r                         % Rank
  l                            % Mode
  S[table-format=1.4]          % Avg Score (4 dp)
  S[table-format=5,            % Total—no decimals
    round-precision=0]@{}}     
\toprule
\textbf{Rank} & \textbf{Mode} & \textbf{Avg Score} & \textbf{Total}\\
\midrule
\rowcolor{yellow!20} 1  & \ourss (comb. extraction)      & 5.3569 & 2657 \\
\rowcolor{yellow!20} 2  & \ourss (comb. embedding)       & 5.2661 & 2612 \\
% \rowcolor{yellow!20} 3  & \ourss (vanilla)\_averaged\_embedding       & 5.2540 & 2606 \\
\rowcolor{yellow!20} 3  & \ourss (hyp. extraction)  & 5.1371 & 2548 \\
4  & LightRAG hybrid                                         & 5.1351 & 2547 \\
5  & \ourss (hyp. embedding)   & 5.1069 & 2533 \\
6  & vanilla RAG                                          & 5.0242 & 2492 \\
7  & GraphRAG drift                                & 4.9758 & 2468 \\
8  & GraphRAG local                                & 4.9657 & 2463 \\
9 & \ourss (vanilla)                                  & 4.9274 & 2444 \\
10 & LightRAG global                                         & 4.8286 & 2395 \\
11 & LightRAG local                                          & 4.7157 & 2339 \\
12 & GraphRAG global                               & 4.5988 & 2281 \\
13 & LightRAG hybrid (comb. extraction)             & 4.3609 & 2163 \\
14 & LightRAG hybrid (hyp. embedding)          & 4.3548 & 2160 \\
15 & LightRAG hybrid (comb. embedding)              & 4.3508 & 2158 \\
16 & vanilla HyDE                                 & 4.3206 & 2143 \\
17 & LightRAG hybrid (hyp. extraction)         & 4.3125 & 2139 \\
18 & RQ-RAG                                          & 2.6472 & 1313 \\
\bottomrule
\end{tabular}
\end{table}

%===================== GPT-4o judge =====================
\begin{table}[htbp]
\centering
\small
\caption{Average scores for \texttt{GPT-4o}}
\label{tab:avg-scores-gpt4o}

\rowcolors{3}{gray!10}{white}
\begin{tabular}{
  @{}r
  l
  S[table-format=1.4]
  S[table-format=5,round-precision=0]@{} }
\toprule
\textbf{Rank} & \textbf{Mode} & \textbf{Avg Score} & \textbf{Total}\\
\midrule
\rowcolor{yellow!20} 1  & LightRAG hybrid                                         & 6.9173 & 3431 \\
\rowcolor{yellow!20} 2  & \ourss (comb. extraction)      & 6.8569 & 3401 \\
\rowcolor{yellow!20} 3  & GraphRAG local                                & 6.8085 & 3377 \\
4  & \ourss (comb. embedding)       & 6.8085 & 3377 \\
% 5  & \ourss (vanilla)\_averaged\_embedding       & 6.7883 & 3367 \\
5  & \ourss (hyp. extraction)  & 6.7823 & 3364 \\
6  & GraphRAG drift                                & 6.7742 & 3360 \\
7  & LightRAG local                                          & 6.6895 & 3318 \\
8  & \ourss (hyp. embedding)   & 6.6895 & 3318 \\
9 & \ourss (vanilla)                                  & 6.6331 & 3290 \\
10 & GraphRAG global                               & 6.5968 & 3272 \\
11 & LightRAG global                                         & 6.5685 & 3258 \\
12 & vanilla RAG                                          & 6.5323 & 3240 \\
13 & LightRAG hybrid (comb. embedding)              & 6.3609 & 3155 \\
14 & LightRAG hybrid (comb. extraction)             & 6.3448 & 3147 \\
15 & LightRAG hybrid (hyp. extraction)         & 6.3407 & 3145 \\
16 & LightRAG hybrid (hyp. embedding)          & 6.3347 & 3142 \\
17 & vanilla HyDE                                 & 6.2581 & 3104 \\
18 & RQ-RAG                                          & 3.9879 & 1978 \\
\bottomrule
\end{tabular}
\end{table}

%===================== GPT-4-1-mini judge =====================
\begin{table}[htbp]
\centering
\small
\caption{Average scores for \texttt{GPT-4.1-mini}}
\label{tab:avg-scores-gpt4mini}

\rowcolors{3}{gray!10}{white}
\begin{tabular}{
  @{}r
  l
  S[table-format=1.4]
  S[table-format=5,round-precision=0]@{} }
\toprule
\textbf{Rank} & \textbf{Mode} & \textbf{Avg Score} & \textbf{Total}\\
\midrule
\rowcolor{yellow!20} 1  & \ourss (comb. extraction)      & 9.1633 & 4545 \\
\rowcolor{yellow!20} 2  & \ourss (comb. embedding)       & 9.1411 & 4534 \\
% \rowcolor{yellow!20} 3  & \ourss (vanilla)\_averaged\_embedding       & 9.1230 & 4525 \\
\rowcolor{yellow!20} 3 & \ourss (hyp. extraction)  & 9.0302 & 4479 \\
4  & \ourss (hyp. embedding)   & 9.0222 & 4475 \\
5  & GraphRAG drift                                & 8.7117 & 4321 \\
6  & GraphRAG local                                & 8.6250 & 4278 \\
7  & LightRAG hybrid                                         & 8.5887 & 4260 \\
8  & \ourss (vanilla)                                  & 8.5645 & 4248 \\
9 & vanilla HyDE                                 & 8.4879 & 4210 \\
10 & GraphRAG global                               & 8.3306 & 4132 \\
11 & vanilla RAG                                          & 8.2500 & 4092 \\
12 & LightRAG local                                          & 8.2440 & 4089 \\
13 & LightRAG global                                         & 7.9778 & 3957 \\
14 & LightRAG hybrid (comb. embedding)              & 6.3790 & 3164 \\
15 & LightRAG hybrid (comb. extraction)             & 6.3387 & 3144 \\
16 & LightRAG hybrid (hyp. embedding)          & 6.3246 & 3137 \\
17 & LightRAG hybrid (hyp. extraction)         & 6.3125 & 3131 \\
18 & RQ-RAG                                          & 3.9073 & 1938 \\
\bottomrule
\end{tabular}
\end{table}

% =========================== Causal Consistency ==========================
\begin{table*}[!b]
\centering
\small
\caption{Causal Consistency category: average scores}
\label{tab:causal-consistency}

\rowcolors{3}{gray!10}{white}

\begin{tabular}{
  @{}r l S[table-format=1.4] S[table-format=5,round-precision=0]@{}}
\toprule
\textbf{Rank} & \textbf{Mode} & \textbf{Avg Score} & \textbf{Total}\\
\midrule
\rowcolor{yellow!20} 1 & \ourss (comb. extraction)      & 7.0590 & 2033 \\
\rowcolor{yellow!20} 2 & \ourss (comb. embedding)       & 6.9896 & 2013 \\
\rowcolor{yellow!20} 3 & \ourss (hyp. extraction)  & 6.8993 & 1987 \\
4  & \ourss (hyp. embedding) & 6.8403 & 1970 \\
5  & \ourss (vanilla)                           & 6.7674 & 1949 \\
6  & GraphRAG drift                         & 6.7535 & 1940 \\
7  & GraphRAG local                         & 6.7326 & 1933 \\
8  & vanilla RAG                            & 6.5694 & 1892 \\
9  & LightRAG hybrid                        & 6.5486 & 1886 \\
10 & LightRAG local                         & 6.4201 & 1859 \\
11 & GraphRAG global                        & 6.3958 & 1842 \\
12 & LightRAG global                        & 6.2674 & 1805 \\
13 & vanilla HyDE                           & 5.7743 & 1663 \\
14 & LightRAG hybrid (comb. extraction)             & 4.9306 & 1411 \\
15 & LightRAG hybrid (hyp. embedding)          & 4.9028 & 1413 \\
16 & LightRAG hybrid (comb. embedding)              & 4.9028 & 1413 \\
17 & LightRAG hybrid (hyp. extraction)         & 4.8993 & 1411 \\
18 & RQ-RAG                                  & 3.1458 &  907 \\
\bottomrule
\end{tabular}
\end{table*}

% ======================= Character Consistency ===========================
\begin{table*}[!b]
\centering
\small
\caption{Character Consistency category: average scores}
\label{tab:character-consistency}

\rowcolors{3}{gray!10}{white}

\begin{tabular}{
  @{}r l S[table-format=1.4] S[table-format=5,round-precision=0]@{}}
\toprule
\textbf{Rank} & \textbf{Mode} & \textbf{Avg Score} & \textbf{Total}\\
\midrule
\rowcolor{yellow!20} 1 & \ourss (comb. extraction)      & 7.2799 & 4994 \\
\rowcolor{yellow!20} 2 & \ourss (comb. embedding)       & 7.2391 & 4966 \\
\rowcolor{yellow!20} 3 & \ourss (hyp. extraction)  & 7.1574 & 4910 \\
4  & \ourss (hyp. embedding) & 7.1429 & 4900 \\
5  & LightRAG hybrid                        & 7.1210 & 4885 \\
6  & GraphRAG drift                         & 6.9752 & 4783 \\
7  & GraphRAG local                         & 6.9621 & 4772 \\
8  & \ourss (vanilla)                           & 6.9096 & 4743 \\
9  & vanilla RAG                            & 6.8192 & 4687 \\
10 & LightRAG local                         & 6.7974 & 4664 \\
11 & GraphRAG global                        & 6.7026 & 4595 \\
12 & vanilla HyDE                           & 6.7012 & 4594 \\
13 & LightRAG global                        & 6.6443 & 4550 \\
14 & LightRAG hybrid (comb. embedding)              & 6.1050 & 4195 \\
15 & LightRAG hybrid (comb. extraction)             & 6.0802 & 4178 \\
16 & LightRAG hybrid (hyp. embedding)          & 6.0700 & 4171 \\
17 & LightRAG hybrid (hyp. extraction)         & 6.0452 & 4153 \\
18 & RQ-RAG                                  & 3.7187 & 2552 \\
\bottomrule
\end{tabular}
\end{table*}

% ================= Emotional & Psychological ============================
\begin{table*}[!b]
\centering
\small
\caption{Emotional and Psychological category: average scores}
\label{tab:emotional-psychological}

\rowcolors{3}{gray!10}{white}

\begin{tabular}{
  @{}r l S[table-format=1.4] S[table-format=4,round-precision=0]@{}}
\toprule
\textbf{Rank} & \textbf{Mode} & \textbf{Avg Score} & \textbf{Total}\\
\midrule
\rowcolor{yellow!20} 1 & \ourss (hyp. extraction)  & 8.3333 &  50 \\
\rowcolor{yellow!20} 2 & LightRAG local                               & 8.1667 &  49 \\
\rowcolor{yellow!20} 3 & \ourss (comb. extraction)      & 8.1667 &  49 \\
4  & \ourss (hyp. embedding) & 8.0000 &  48 \\
5  & vanilla RAG                            & 8.0000 &  48 \\
6  & LightRAG hybrid                        & 7.8333 &  47 \\
7  & \ourss (comb. embedding)  & 7.8333 &  47 \\
8  & \ourss (vanilla)                           & 7.6667 &  46 \\
9  & LightRAG hybrid (hyp. embedding)          & 7.1667 &  43 \\
10 & LightRAG hybrid (comb. extraction)             & 7.1667 &  43 \\
11 & LightRAG global                        & 7.1667 &  43 \\
12 & LightRAG hybrid (hyp. extraction)         & 7.1667 &  43 \\
13 & LightRAG hybrid (comb. embedding)              & 7.1667 &  43 \\
14 & GraphRAG drift                         & 7.0000 &  42 \\
15 & vanilla HyDE                           & 6.6667 &  40 \\
16 & GraphRAG local                         & 6.3333 &  38 \\
17 & GraphRAG global                        & 6.3333 &  38 \\
18 & RQ-RAG                                  & 6.0000 &  36 \\
\bottomrule
\end{tabular}
\end{table*}

% ================= Narrative & Plot Structure ============================
\begin{table*}[!b]
\centering
\small
\caption{Narrative and Plot Structure category: average scores}
\label{tab:narrative-plot}

\rowcolors{3}{gray!10}{white}

\begin{tabular}{
  @{}r l S[table-format=1.4] S[table-format=4,round-precision=0]@{}}
\toprule
\textbf{Rank} & \textbf{Mode} & \textbf{Avg Score} & \textbf{Total}\\
\midrule
\rowcolor{yellow!20} 1 & \ourss (hyp. extraction)  & 6.9785 & 649 \\
\rowcolor{yellow!20} 2 & \ourss (comb. extraction)      & 6.9462 & 646 \\
\rowcolor{yellow!20} 3 & GraphRAG drift                               & 6.8817 & 640 \\
4  & \ourss (comb. embedding)  & 6.8280 & 635 \\
5  & \ourss (hyp. embedding) & 6.7204 & 625 \\
6  & LightRAG hybrid                        & 6.6882 & 622 \\
7  & GraphRAG local                         & 6.6774 & 621 \\
8  & \ourss (vanilla)                           & 6.6129 & 615 \\
9  & LightRAG local                         & 6.4624 & 601 \\
10 & LightRAG global                        & 6.4194 & 597 \\
11 & vanilla RAG                            & 6.3011 & 586 \\
12 & GraphRAG global                        & 6.2903 & 585 \\
13 & vanilla HyDE                           & 6.0753 & 565 \\
14 & LightRAG hybrid (hyp. embedding)          & 5.7204 & 532 \\
15 & LightRAG hybrid (comb. extraction)             & 5.7204 & 532 \\
16 & LightRAG hybrid (comb. embedding)              & 5.7204 & 532 \\
17 & LightRAG hybrid (hyp. extraction)         & 5.7097 & 531 \\
18 & RQ-RAG                                  & 3.6129 & 336 \\
\bottomrule
\end{tabular}
\end{table*}

% =========== Setting, Environment & Atmosphere ===========================
\begin{table*}[!b]
\centering
\small
\caption{Setting, Environment and Atmosphere category: average scores}
\label{tab:setting-environment}

\rowcolors{3}{gray!10}{white}

\begin{tabular}{
  @{}r l S[table-format=1.4] S[table-format=4,round-precision=0]@{}}
\toprule
\textbf{Rank} & \textbf{Mode} & \textbf{Avg Score} & \textbf{Total}\\
\midrule
\rowcolor{yellow!20} 1 & \ourss (comb. extraction)      & 6.4795 & 473 \\
\rowcolor{yellow!20} 2 & \ourss (comb. embedding)       & 6.2877 & 459 \\
\rowcolor{yellow!20} 3 & LightRAG hybrid                              & 6.2603 & 457 \\
4  & \ourss (hyp. extraction) & 6.2603 & 457 \\
5  & vanilla RAG                            & 6.1507 & 449 \\
6  & \ourss (hyp. embedding) & 6.1233 & 447 \\
7  & \ourss (vanilla)                           & 6.0000 & 438 \\
8  & GraphRAG drift                         & 5.9452 & 434 \\
9  & LightRAG local                         & 5.8356 & 426 \\
10 & vanilla HyDE                           & 5.7534 & 420 \\
11 & GraphRAG local                         & 5.6438 & 412 \\
12 & GraphRAG global                        & 5.5890 & 409 \\
13 & LightRAG global                        & 5.5479 & 404 \\
14 & LightRAG hybrid (hyp. embedding)          & 4.7397 & 346 \\
15 & LightRAG hybrid (comb. embedding)              & 4.7397 & 346 \\
16 & LightRAG hybrid (comb. extraction)             & 4.7260 & 345 \\
17 & LightRAG hybrid (hyp. extraction)         & 4.7260 & 345 \\
18 & RQ-RAG                                  & 3.0137 & 220 \\
\bottomrule
\end{tabular}
\end{table*}

% ============ Social, Cultural & Political ==============================
\begin{table*}[!b]
\centering
\small
\caption{Social, Cultural and Political category: average scores}
\label{tab:social-cultural}

\rowcolors{3}{gray!10}{white}

\begin{tabular}{
  @{}r l S[table-format=1.4] S[table-format=4,round-precision=0]@{}}
\toprule
\textbf{Rank} & \textbf{Mode} & \textbf{Avg Score} & \textbf{Total}\\
\midrule
\rowcolor{yellow!20} 1 & \ourss (comb. extraction)      & 7.0758 & 467 \\
\rowcolor{yellow!20} 2 & \ourss (comb. embedding)       & 6.9697 & 460 \\
\rowcolor{yellow!20} 3 & LightRAG hybrid                              & 6.9242 & 457 \\
4  & \ourss (hyp. extraction) & 6.8636 & 453 \\
5  & \ourss (hyp. embedding) & 6.8333 & 451 \\
6  & GraphRAG local                         & 6.6364 & 438 \\
7  & LightRAG global                        & 6.6212 & 437 \\
8  & GraphRAG drift                         & 6.5909 & 435 \\
9  & GraphRAG global                        & 6.5606 & 433 \\
10 & vanilla HyDE                           & 6.5303 & 431 \\
11 & vanilla RAG                            & 6.4848 & 428 \\
12 & \ourss (vanilla)                           & 6.4545 & 426 \\
13 & LightRAG local                         & 6.3333 & 418 \\
14 & LightRAG hybrid (comb. embedding)              & 5.8788 & 389 \\
15 & LightRAG hybrid (hyp. embedding)          & 5.8182 & 384 \\
16 & LightRAG hybrid (comb. extraction)             & 5.8030 & 383 \\
17 & LightRAG hybrid (hyp. extraction)         & 5.7879 & 382 \\
18 & RQ-RAG                                  & 3.2121 & 212 \\
\bottomrule
\end{tabular}
\end{table*}

% ============ Symbolism, Imagery & Motifs ===============================
\begin{table*}[!b]
\centering
\small
\caption{Symbolism, Imagery and Motifs category: average scores}
\label{tab:symbolism}

\rowcolors{3}{gray!10}{white}

\begin{tabular}{
  @{}r l S[table-format=1.4] S[table-format=5,round-precision=0]@{}}
\toprule
\textbf{Rank} & \textbf{Mode} & \textbf{Avg Score} & \textbf{Total}\\
\midrule
\rowcolor{yellow!20} 1 & \ourss (comb. embedding)       & 6.8393 & 1149 \\
\rowcolor{yellow!20} 2 & \ourss (comb. extraction)      & 6.8214 & 1146 \\
\rowcolor{yellow!20} 3 & \ourss (hyp. extraction)  & 6.6845 & 1123 \\
4  & GraphRAG local                         & 6.6548 & 1118 \\
5  & \ourss (hyp. embedding) & 6.6488 & 1117 \\
6  & LightRAG hybrid                        & 6.6190 & 1112 \\
7  & GraphRAG drift                         & 6.5000 & 1092 \\
8  & \ourss (vanilla)                           & 6.2500 & 1050 \\
9  & GraphRAG global                        & 6.2024 & 1042 \\
10 & LightRAG global                        & 6.1667 & 1037 \\
11 & vanilla RAG                            & 6.1310 & 1020 \\
12 & LightRAG local                         & 6.1131 & 1022 \\
13 & vanilla HyDE                           & 6.0714 & 1020 \\
14 & LightRAG hybrid (comb. embedding)              & 5.3393 &  898 \\
15 & LightRAG hybrid (hyp. embedding)          & 5.3214 &  894 \\
16 & LightRAG hybrid (comb. extraction)             & 5.3214 &  894 \\
17 & LightRAG hybrid (hyp. extraction)         & 5.3155 &  894 \\
18 & RQ-RAG                                  & 3.4464 &  580 \\
\bottomrule
\end{tabular}
\end{table*}

% ============ Thematic, Philosophical & Moral ============================
\begin{table*}[!b]
\centering
\small
\caption{Thematic, Philosophical and Moral category: average scores}
\label{tab:thematic}

\rowcolors{3}{gray!10}{white}

\begin{tabular}{
  @{}r l S[table-format=1.4] S[table-format=4,round-precision=0]@{}}
\toprule
\textbf{Rank} & \textbf{Mode} & \textbf{Avg Score} & \textbf{Total}\\
\midrule
\rowcolor{yellow!20} 1 & \ourss (comb. extraction)      & 7.3611 & 795 \\
\rowcolor{yellow!20} 2 & \ourss (comb. embedding)       & 7.3519 & 794 \\
\rowcolor{yellow!20} 3 & GraphRAG local                               & 7.1852 & 776 \\
4  & GraphRAG drift                         & 7.1852 & 776 \\
5  & LightRAG hybrid                        & 7.1481 & 772 \\
6  & \ourss (hyp. embedding) & 7.1111 & 768 \\
7  & \ourss (hyp. extraction) & 7.0556 & 762 \\
8  & GraphRAG global                        & 6.8426 & 738 \\
9  & LightRAG global                        & 6.7500 & 729 \\
10 & vanilla HyDE                           & 6.6759 & 721 \\
11 & \ourss (vanilla)                           & 6.6481 & 718 \\
12 & vanilla RAG                            & 6.6019 & 713 \\
13 & LightRAG local                         & 6.6019 & 713 \\
14 & LightRAG hybrid (comb. embedding)              & 6.2130 & 671 \\
15 & LightRAG hybrid (comb. extraction)             & 6.1667 & 667 \\
16 & LightRAG hybrid (hyp. embedding)          & 6.1481 & 664 \\
17 & LightRAG hybrid (hyp. extraction)         & 6.1389 & 664 \\
18 & RQ-RAG                                  & 3.6019 & 389 \\
\bottomrule
\end{tabular}
\end{table*}

\FloatBarrier
\section{Pseudocode}
\label{subsec:Pseudocode}

\begin{algorithm}
\caption{Entity–Event KG Insertion (Pre-processing)}
\label{alg:insertion}
\begin{algorithmic}[1]
\Require
  document $D$;  
  entity, event extractor $g(\cdot)$;  
 
  text encoder   $f(\cdot)$;  
  top-$L$ chunk size $m$
\Ensure
  graphs \(\mathcal{G}_{\text{ent}},\mathcal{G}_{\text{evt}}\);  
  bipartite edge set \(\mathcal{B}\);  
  vector stores \(\mathcal{I}_{\mathrm{ent}},\mathcal{I}_{\mathrm{evt}}\)
\Procedure{InsertDocument}{$D$}
  %-----------------------------------------------------------
  \State \textbf{Step 1: chunking}
  \State $\{c_\ell\}_{\ell=1}^{L}\gets\textsc{Chunk}(D,m)$
  %-----------------------------------------------------------
  \State \textbf{Step 2: node extraction}
  \For{$\ell\gets1$ \textbf{to} $L$}
    \State $V_{\mathrm{ent}}^{(\ell)}, V_{\mathrm{evt}}^{(\ell)}\gets g(c_\ell)$
  \EndFor
  \State $V_{\mathrm{ent}}\gets\bigcup_\ell V_{\mathrm{ent}}^{(\ell)}$;\;
        $V_{\mathrm{evt}}\gets\bigcup_\ell V_{\mathrm{evt}}^{(\ell)}$
  %-----------------------------------------------------------
  \State \textbf{Step 3: build edge sets}
  \State $E_{\mathrm{ent}},E_{\mathrm{evt}}\gets
         \textsc{IntraChunkEdges}(\{c_\ell\},V_{\mathrm{ent}},V_{\mathrm{evt}})$
  \State $\mathcal{B}\gets\Call{BuildBipartite}{V_{\mathrm{ent}},V_{\mathrm{evt}}}$
  %-----------------------------------------------------------
  \State \textbf{Step 4: embed and index}
  \ForAll{$v\in V_{\mathrm{ent}}$}  \Comment{entities}
      \State $\mathcal{I}_{\mathrm{ent}}.\textsc{Add}\bigl(v,f(\textsc{Canon}(v))\bigr)$
  \EndFor
  \ForAll{$e\in V_{\mathrm{evt}}$}  \Comment{events}
      \State $\mathcal{I}_{\mathrm{evt}}.\textsc{Add}\bigl(e,f(\textsc{Canon}(e))\bigr)$
  \EndFor
  %-----------------------------------------------------------
  \State \Return \(\mathcal{G}_{\text{ent}}=(V_{\mathrm{ent}},E_{\mathrm{ent}}),
                  \;\mathcal{G}_{\text{evt}}=(V_{\mathrm{evt}},E_{\mathrm{evt}}),
                  \;\mathcal{B}\)
\EndProcedure
\end{algorithmic}
\end{algorithm}

\begin{algorithm}
\caption{\textsc{BuildBipartite}: connect entity mentions to events}
\label{alg:bipartite}
\begin{algorithmic}[1]
\Require
  entity nodes $V_{\mathrm{ent}}$,  
  event nodes  $V_{\mathrm{evt}}$
\Ensure
  bipartite edge set \(\mathcal{B}\)
\Function{BuildBipartite}{$V_{\mathrm{ent}},V_{\mathrm{evt}}$}
  \State $\mathcal{B}\gets\varnothing$
  \State \textit{/* surface-form string match inside the same chunk */}  
  \ForAll{$v\in V_{\mathrm{ent}}$}
    \State $c\gets\textsc{ChunkID}(v)$;\;
           $S\gets\textsc{Name}(v)$
    \ForAll{$e\in V_{\mathrm{evt}}\;\textbf{with}\;\textsc{ChunkID}(e)=c$}
        \If{$S\subset\textsc{Desc}(e)$}
            \State $\mathcal{B}\gets\mathcal{B}\cup\{(v,e)\}$
        \EndIf
    \EndFor
  \EndFor
  \State \Return $\mathcal{B}$
\EndFunction
\end{algorithmic}
\end{algorithm}

% \begin{algorithm}
% \caption{Entity-KG builder (wrapper)}
% \begin{algorithmic}
% \Function{ConstructEntityKG}{$D$}
%   \State \Return $\Call{InsertDocument}{D}.\mathcal{G}_{\text{ent}}$
% \EndFunction
% \end{algorithmic}
% \end{algorithm}

\begin{algorithm}
\caption{Entity–Event KG \textsc{Retrieve} (Inference-time)}
\label{alg:retrieval}
\begin{algorithmic}[1]
\Require
  query $q$ (or hypothetical response transformed text),\\
  entity store $\mathcal{I}_{\mathrm{ent}}$, event store $\mathcal{I}_{\mathrm{evt}}$,\\
  bipartite edge set $\mathcal{B}$, similarity encoder $f(\cdot)$,\\
  entity, event extractor $g(\cdot)$,   top-$k$ parameter $k$
\Ensure
  ranked context bundle $C$ (passages $+$ linearised subgraph)
\Procedure{Retrieve}{$q$}
  %-----------------------------------------------------------
  \State \textbf{Cue extraction}
  \State $S_{\mathrm{ent}},S_{\mathrm{evt}}\gets g(q)$
  \State $V_q \gets \textsc{Lookup}(S_{\mathrm{ent}},\mathcal{I}_{\mathrm{ent}})
                 \cup
                 \textsc{Lookup}(S_{\mathrm{evt}},\mathcal{I}_{\mathrm{evt}})$
  %-----------------------------------------------------------
  \State \textbf{One-hop expansion across $\mathcal{B}$}
  \State $V_q^{+}\!\gets\!V_q
        \cup\{e\mid(v,e)\!\in\!\mathcal{B},\,v\!\in\!V_q\}
        \cup\{v\mid(v,e)\!\in\!\mathcal{B},\,e\!\in\!V_q\}$
  %-----------------------------------------------------------
  \State \textbf{Similarity ranking}
  \State $\mathcal{P}\gets\{\textsc{Chunk}(x)\mid x\in V_q^{+}\}$ \Comment{candidate passages}
  \State $\mathbf{z}\gets f(q)$
  \State $\text{scores}[p]\gets\langle f(p),\mathbf{z}\rangle\quad\forall p\in\mathcal{P}$
  \State $\mathcal{P}_k\gets\textsc{TopK}(\text{scores},k)$
  %-----------------------------------------------------------
  \State \textbf{Context assembly}
  \State $G_{\text{sub}}\gets
         (\,V_q^{+},\,
           (E_{\mathrm{ent}}\cup E_{\mathrm{evt}}\cup\mathcal{B})
           \cap (V_q^{+}\!\times\!V_q^{+})\,)$
  \State $C\gets\textsc{FormatContext}(\mathcal{P}_k,G_{\text{sub}})$
  \State \Return $C$
\EndProcedure
\end{algorithmic}
\end{algorithm}

\end{document}

%% file: notations.tex
\usepackage{xspace}
\usepackage{bm}

\newcommand{\benchmark}{ChronoQA}
\newcommand{\ours}{$\mathtt{E^2RAG}$\xspace}
\newcommand{\ourss}{\texttt{\textbf{E\textsuperscript{2}RAG}}\xspace}